\RequirePackage{fixltx2e}
\documentclass[11pt,a4paper]{article}
\pdfoutput=1
\usepackage[utf8]{inputenc}
\usepackage{jcappub}
\hypersetup{unicode=true,bookmarksopen=true}

\usepackage{mathtools}
\usepackage{lmodern}
\usepackage[T1]{fontenc}
\usepackage{bbm}
\DeclareMathAlphabet{\mathbfi}{OML}{cmm}{b}{it}

\renewcommand{\vec}[1]{{\ifnum9<1#1\mathbf{#1}\else\ifcat\noexpand#1\relax\boldsymbol{#1}\else\mathbfi{#1}\fi\fi}}
\newcommand{\mathe}{\mathrm{e}}
\newcommand{\mathi}{\mathrm{i}}
\let\oldre\Re
\let\oldim\Im
\renewcommand{\Re}{\oldre\mathfrak{e}\,}
\renewcommand{\Im}{\oldim\mathfrak{m}\,}
\newcommand{\total}{\mathop{}\!\mathrm{d}}
\newcommand{\laplace}{\mathop{}\!\bigtriangleup}
\newcommand{\abs}[1]{{\left\lvert{#1}\right\rvert}}
\newcommand{\eqend}[1]{\,#1}
\newcommand{\bigo}[1]{\mathcal{O}\!\left({#1}\right)}
\newcommand{\Ein}{\operatorname{Ein}}
\DeclareMathOperator*{\distlim}{d-lim}
\newcommand{\bra}[1]{\left\langle{#1}\right\vert}
\newcommand{\ket}[1]{\left\vert{#1}\right\rangle}

\allowdisplaybreaks

\begin{document}

\title{Quantum corrections to the gravitational potentials of a point source due to conformal fields in de~Sitter}

\author[1]{Markus B. Fröb}
\affiliation[1]{Institut für Theoretische Physik, Universität Leipzig,\\ Brüderstraße 16, 04103 Leipzig, Germany}

\author[2]{and Enric Verdaguer}
\affiliation[2]{Departament de Física Fonamental, Institut de Ciències del Cosmos (ICC), Universitat de\\ Barcelona (UB), C/ Martí i Franquès 1, 08028 Barcelona, Spain}

\emailAdd{mfroeb@itp.uni-leipzig.de}

\emailAdd{enric.verdaguer@ub.edu}

\abstract{We derive the leading quantum corrections to the gravitational potentials in a de~Sitter background, due to the vacuum polarization from loops of conformal fields. Our results are valid for arbitrary conformal theories, even strongly interacting ones, and are expressed using the coefficients $b$ and $b'$ appearing in the trace anomaly. Apart from the de~Sitter generalization of the known flat-space results, we find two additional contributions: one which depends on the finite coefficients of terms quadratic in the curvature appearing in the renormalized effective action, and one which grows logarithmically with physical distance. While the first contribution corresponds to a rescaling of the effective mass, the second contribution leads to a slower fall-off of the Newton potential at large distances, and is potentially measurable.}


\maketitle

\section{Introduction}

The computation of the leading quantum corrections to the Newtonian gravitational potential is an old topic that was already discussed by Radkowski and Schwinger half a century ago~\cite{radkowski1970,schwinger1968}. These calculations have been used as a testing ground for the effective field theory description of quantum gravity~\cite{donoghue1994a,donoghue1994b}, in the sense that such corrections are necessary consequences of any theory of quantum gravity, independently of its high-energy completion. More recently~\cite{duffliu2000a,duffliu2000b,mueckviswanathanvolovich2000,garrigatanaka2000}, these quantum corrections have also been used to probe the pictures that emerge from the AdS/CFT correspondence~\cite{maldacena1998,witten1998} and the Randall-Sundrum braneworld scenario~\cite{randallsundrum1999}. The quantum-corrected Newtonian potential including loops of different fields is by now known for free scalar, spinor and vector fields~\cite{duff1974,capperduffhalpern1974,capperduff1974,duffliu2000a,duffliu2000b,parkwoodard2010}, and for the more complicated case of graviton fields~\cite{donoghue1994a,donoghue1994b,muzinichvokos1995,hamberliu1995,akhundovbelluccishiekh1997,kirilinkhriplovich2002,khriplovichkirilin2003,bjerrumbohrdonoghueholstein2003a,bjerrumbohrdonoghueholstein2003b}. In all these cases, the correction to the $1/r$ Newtonian potential is given by terms that go like $\ell_\text{Pl}^2/r^3$, where $\ell_\text{Pl}$ is the Planck length, with a coefficient that depends on the type of matter loop considered.

One way to compute these quantum corrections is to follow the path taken for the quantum corrections to the Coulomb potential, which are usually computed from the non-relativistic limit of the scattering amplitude of two charged electrons~\cite{peskinschroeder}. In a similar way, the leading quantum corrections to the Newtonian potential may be inferred from the non-relativistic limit of the scattering amplitude of two heavy particles at rest by graviton exchange~\cite{radkowski1970,donoghue1994a,donoghue1994b}. An alternative way, however, is to derive these leading corrections from the effective field equations for the linearized metric perturbations that couple to a given field. The procedure in this case is to compute first the graviton self-energy up to one-loop order in the given field, which is needed for the effective field equations, and then compute the gravitational response to a massive point source~\cite{duff1974,duffliu2000a,duffliu2000b,parkwoodard2010}. As Park and Woodard have recently pointed out~\cite{parkwoodard2010}, this method not only emphasizes the similarity with the classical calculation and allows for the possibility of studying the response to dynamical sources, but is also much easier in practice. Furthermore, it can also be used on non-trivial curved backgrounds such as cosmological ones including de~Sitter space, where scattering amplitudes are not even well defined~\cite{bousso2005}. However, to obtain real and causal effective field equations one should calculate using the Schwinger-Keldysh or in-in formalism~\cite{schwinger1961,keldysh1964,chousuhaoyu1985}, which is also necessary to avoid problems with the in-out formalism that appear in curved spacetimes~\cite{polyakov2007,higuchi2008,higuchilee2009}.

In this work, we compute the leading corrections to the gravitational potentials on a de~Sitter background, with the quantum effects coming from loops of conformal matter (such as the massless, conformally coupled scalar, massless fermions, or photons). This is done in accordance with the second method described, fixing a static point mass source in the Poincar\'e patch of de~Sitter space and solving the effective field equations for linearized metric perturbations coupling both to the quantum matter and the point mass. These equations are derived from an ``in-in'' or closed-time-path (CTP) renormalized effective action, which guarantees both causal equations and the proper renormalization of the usual field theoretic UV divergences, such that no further renormalization needs to be done in the field equations. To suppress the contribution from graviton loops (i.e., self-interaction of the metric perturbations) which are difficult to handle consistently both from a theoretical and a calculational viewpoint, but which in general arise in the effective action, we work to leading order in a $1/N$-expansion, where $N$ is the number of conformal fields that couple to the gravitational field. We furthermore neglect the backreaction of the point mass on the de~Sitter background, which would give contributions comparable with the ones from graviton loops. A useful technical device is the decomposition of the metric perturbations according to their transformation properties under rotations and translations on the spatially flat slices into scalar, vector and tensor perturbations, which are independent to the order that we are interested in. The static point mass, our source to probe the Newtonian potential, only couples to one of the scalar perturbations and we can thus ignore vectors and tensors, simplifying our calculation. A second simplification comes about by expressing the effective action using the gauge-invariant Bardeen potentials~\cite{bardeen1980}, which halves the number of independent fields since the gauge-dependent parts of the metric perturbation drop out.

There is interest in doing this calculation on a de~Sitter background, because de~Sitter spacetime plays a very important role in cosmology. Even if we expect corrections that are at present too small to be observable, it is important to test perturbative quantum field theory and its interaction with metric perturbations, both quantized and classical, beyond tree level. Quantum corrections to the gravitational potentials on de~Sitter have also been studied by Wand and Woodard~\cite{wangwoodard2015} for the case of photons, and by Park, Prokopec and Woodard~\cite{parkprokopecwoodard2015} for massless, minimally coupled scalars, and their results and the present calculation are steps in this direction.

The paper is organized as follows: in Sec.~\ref{sec_effectiveaction}, we derive the effective action for the coupled quantum system point mass -- metric perturbations -- conformal fields, perform the decomposition of the metric perturbations and express the effective action using the two Bardeen potentials. In Sec.~\ref{sec_fieldequations}, we derive the effective field equations for these potentials, and show the classical and quantum contributions to these equations. In Sec.~\ref{sec_solutions}, we proceed to solve the equations, study the flat-space limit and compare with previous works. Sec.~\ref{sec_discussion} comprises the discussion, while a lengthy calculation of a non-local integral and some technical parts are delegated to the appendices. We use the ``+++'' convention of Ref.~\cite{mtw}, units such that $c = \hbar = 1$, and define $\kappa^2 \equiv 16 \pi G_\text{N}$ with Newton's constant $G_\text{N}$. Greek indices range over spacetime, while Latin ones are purely spatial.

\section{Effective action}
\label{sec_effectiveaction}

To study the effect of conformal fields on the Newtonian potential in a de~Sitter background, we solve the effective field equations which generalize the Poisson equation $\laplace \phi(\vec{x}) = 4 \pi G_\text{N} m \delta^3(\vec{x})$ for a point source of mass $m$, where $\laplace$ is the flat space Laplacian. These equations can be derived from an effective action $S_\text{eff}$, which includes the quantum radiative corrections due to loops of conformal fields, and which can be obtained by integrating out these matter fields from the bare action $S$. We thus consider a bare action of the form
\begin{equation}
\label{bare_action}
S = S_\text{G} + S_\text{M} + S_\text{CT} + S_\text{PP} \eqend{,}
\end{equation}
where
\begin{equation}
\label{action_eh}
S_\text{G} = \frac{1}{\kappa^2} \int \left( \tilde{R} - 2 \Lambda \right) \sqrt{-\tilde{g}} \total^n x
\end{equation}
is the Einstein-Hilbert gravitational action for a metric $\tilde{g}_{\mu\nu}$ including a positive cosmological constant $\Lambda$,
\begin{equation}
\label{action_pp}
S_\text{PP} = \frac{m}{2} \int \tilde{g}_{\mu\nu}(x) \int \frac{\total z^\mu(\tau)}{\total \tau} \frac{\total z^\nu(\tau)}{\total \tau} \delta^4(x-z(\tau)) \total \tau \total^4 x
\end{equation}
is the action for a point particle of mass $m$ with four-position $z^\mu(\tau)$ and normalized four-velocity $\tilde{g}_{\mu\nu}(z(\tau)) (\total z^\mu/\total \tau) (\total z^\nu/\total \tau) = -1$, $S_\text{M}$ is the action for conformal matter, e.g., free massless, conformally coupled scalar fields $\tilde\phi$
\begin{equation}
\label{s_matter_ccs}
S_\text{M} = - \frac{1}{2} \int \left[ \left( \tilde{\nabla}^\mu \tilde{\phi} \right) \left( \tilde{\nabla}_\mu \tilde{\phi} \right) + \xi(n) \tilde{R} \tilde{\phi}^2 \right] \sqrt{-\tilde{g}} \total^n x
\end{equation}
with the conformal coupling $\xi(n) = (n-2)/[4(n-1)]$, and $S_\text{CT}$ are counterterms quadratic in the curvature tensors which are needed for renormalization. We use dimensional regularization and renormalization, and correspondingly our bare action is taken in $n$ dimensions.

\subsection{The in-in formalism}

To obtain real and causal field equations, one has to use the Schwinger-Keldysh/closed-time-path/in-in formalism~\cite{schwinger1961,keldysh1964,chousuhaoyu1985}, instead of the usual flat-space in-out formalism. In the de~Sitter spacetime, the \emph{in} and \emph{out} states are orthogonal to each other for an arbitrary small interaction~\cite{polyakov2007,higuchi2008}, and it has been emphasized that in-out perturbation theory is thus inadequate since it presupposes at least some overlap between these vacuum states. This may be illustrated by a free scalar field theory with mass $\sqrt{m_1^2 + m_2^2}$, treating $m_2^2$ as a perturbation of the theory with mass $m_1$. When resumming all diagrams, in-out perturbation theory gives completely wrong results, while the in-in theory recovers the true (free) theory with the correct mass $\sqrt{m_1^2 + m_2^2}$~\cite{higuchilee2009}. (For a discussion of the necessity of the in-in formalism in the context of corrections to the Newtonian potential, see~\cite{parkwoodard2010}.) In the in-in formalism, one doubles the number of fields (which are distinguished by a subscript ``$+$''/``$-$''), where the ``$+$'' fields correspond to the usual in-out formalism, while the ``$-$'' fields enter the path integral with the complex conjugate action. One then enforces equality of both fields at some final time $T$ that must be larger than all times of interest (and may be sent to future infinity), and integrates over all types of fields.

The effective action $S_\text{eff}$ results from functionally integrating over the matter fields
\begin{equation}
\label{effective_action}
\mathe^{\mathi S_\text{eff}[ \tilde{g}^\pm ]} \equiv \int \mathe^{\mathi S[\tilde{g}^+,\tilde{\phi}^+] - \mathi S[\tilde{g}^-,\tilde{\phi}^-]} \delta(\tilde{\phi}^+(T) - \tilde{\phi}^-(T)) \mathcal{D} \tilde{\phi}^\pm \eqend{,}
\end{equation}
where $S$ is the full bare action of Eq.~\eqref{bare_action}. Defining
\begin{equation}
\label{sloc_def}
S_\text{loc} \equiv S_\text{G} + S_\text{PP} + S_\text{CT} \eqend{,}
\end{equation}
which is the part of the bare action $S$ that only depends on the metric, and is thus not affected by the integration over the matter fields, we obtain
\begin{equation}
\label{seff_funcint}
\mathe^{\mathi S_\text{eff}[ \tilde{g}^\pm ]} = \mathe^{\mathi S_\text{loc}[ \tilde{g}^+ ] - \mathi S_\text{loc}[ \tilde{g}^- ]} \int \mathe^{\mathi S_\text{M}[\tilde{g}^+,\tilde{\phi}^+] - \mathi S_\text{M}[\tilde{g}^-,\tilde{\phi}^-]} \delta(\tilde{\phi}^+(T) - \tilde{\phi}^-(T)) \mathcal{D} \tilde{\phi}^\pm \eqend{.}
\end{equation}
The functional integral now gives a non-local term that we write in the form $\mathe^{\mathi \Sigma\left[ \tilde{g}^\pm \right]}$, such that the effective action has the form
\begin{equation}
\label{seff_in_sloc_sigma}
S_\text{eff}\!\left[ \tilde{g}^\pm \right] = S_\text{loc}\!\left[ \tilde{g}^+ \right] - S_\text{loc}\!\left[ \tilde{g}^- \right] + \Sigma\!\left[ \tilde{g}^\pm \right] \eqend{.}
\end{equation}
Since $\Sigma$ is divergent as $n \to 4$, the counterterms in $S_\text{CT}$ must be chosen to cancel these divergences in order to obtain a finite result for $S_\text{eff}$ in the limit $n \to 4$.

For free matter fields, the above amounts to a one-loop calculation in a fixed gravitational background. As explained in the introduction, in addition to the quantum correction to the Newtonian potential arising from matter fields, there is also a contribution from gravitons. To obtain this contribution from an effective action, one would have to split the metric $\tilde{g}_{\mu\nu}$ into a background metric and perturbations, and then functionally integrate also over the metric perturbations. However, we only want to consider quantum corrections due to conformal matter, and thus must suppress the contribution from graviton loops. This can be consistently done in a $1/N$-expansion, where one considers $N$ matter fields and rescales the gravitational constant $\kappa^2 \to \bar{\kappa}^2 \equiv \kappa^2/N$ with $\bar{\kappa}$ held fixed. For interacting matter fields, one also has to rescale the coupling to obtain a well-defined $1/N$-expansion, e.g., holding $\bar{\lambda} \equiv N \lambda$ fixed for a $\lambda (\phi^2)^2$ interaction in an $\mathrm{O}(N)$ model~\cite{schnitzer1974}. While both the matter action $S_\text{M}$ including all matter interaction vertices and the tree-level part of the purely gravitational action $S_\text{loc}$ then scale proportional to $N$, graviton-graviton interaction vertices (and thus the contribution from graviton loops) are suppressed by powers of $1/N$. In the limit $N \to \infty$, the functional integral can thus be performed in a saddle-point approximation, which includes tree-level terms for the gravitons and all matter loops. In order not to overload the formulas, we will not do this explicitly here but simply restrict to the tree-level terms for the metric perturbations, which are of second order (see Refs.~\cite{frv2012,frv2014} for more details).

In general, while powerful methods are known to calculate the necessary counterterms $S_\text{CT}$ (i.e., the divergent part of $\Sigma$) for arbitrary backgrounds $\tilde{g}$ (such as the heat kernel expansion~\cite{vassilevich2003}), the calculation of the finite parts of $\Sigma$ is a hard task. Even for free matter fields, in general it is not possible to find explicit expressions. One important exception is the case of conformal matter in a conformally flat background (including metric perturbations), i.e., where the metric is of the form
\begin{equation}
\label{conf_metric}
\tilde{g}_{\mu\nu} \equiv a^2 g_{\mu\nu} \equiv a^2 \left( \eta_{\mu\nu} + h_{\mu\nu} \right) \eqend{,}
\end{equation}
where $a(x^\mu)$ is a conformal factor (depending, in general, on all coordinates), $\eta_{\mu\nu}$ is the flat Minkowski metric and $h_{\mu\nu}$ is the metric perturbation. It is then possible to calculate the effective action $S_\text{eff}$~\eqref{seff_in_sloc_sigma} to an arbitrary order in the perturbation $h_{\mu\nu}$, but we restrict to second order and neglect graviton loops as explained in the last paragraph.

\subsection{Conformal theories in conformally flat backgrounds}

For the metric~\eqref{conf_metric}, the explicit calculation of $\Sigma$ has been done by Campos and Verdaguer~\cite{camposverdaguer1994,camposverdaguer1996} for the case of a massless, conformally coupled scalar~\eqref{s_matter_ccs}, and the necessary counterterms are given by
\begin{equation}
S^\text{CV}_\text{CT}[\tilde{g}] = \frac{\bar{\mu}^{n-4}}{2880 \pi^2 (n-4)} \int \left( \tilde{R}^{\alpha\beta\gamma\delta} \tilde{R}_{\alpha\beta\gamma\delta} - \tilde{R}^{\alpha\beta} \tilde{R}_{\alpha\beta} \right) \sqrt{-\tilde{g}} \total^n x \eqend{,}
\end{equation}
where $\bar{\mu}$ is the renormalization scale; the cosmological constant, the Einstein-Hilbert term and the square of the Ricci scalar do not get renormalized. In fact, as also explained in Refs.~\cite{fprv2013,frv2014}, the result for a general conformal theory is essentially the same. In general, the counterterms are given in the minimal subtraction (MS) scheme by~\cite{duff1977}
\begin{equation}
\label{ms_scheme}
S^\text{MS}_\text{CT}[ \tilde{g} ] = \frac{\bar{\mu}^{n-4}}{n-4} \int \left( b \tilde{C}^2 + b' \tilde{\mathcal{E}}_4 \right) \sqrt{-\tilde{g}} \total^n x \eqend{,}
\end{equation}
where $\tilde{C}^2 \equiv \tilde{C}^{\alpha\beta\gamma\delta} \tilde{C}_{\alpha\beta\gamma\delta}$ with the $n$-dimensional Weyl tensor $\tilde{C}_{\alpha\beta\gamma\delta}$ defined in equation~\eqref{weyl_tensor_def}, and where $\tilde{\mathcal{E}}_4$ is the four-dimensional Euler density defined in equation~\eqref{euler_density_def}. The constants $b$ and $b'$ are the exact same coefficients that appear in the trace anomaly in front of the square of the Weyl tensor and the Euler density, and depend on the theory under consideration. For $N_0$ free massless, conformally coupled scalar fields, $N_{1/2}$ free, massless Dirac spinor fields and $N_1$ free vector fields we have~\cite{duff1977}
\begin{subequations}
\label{coeffs_b}
\begin{align}
b &= \frac{N_0 + 6 N_{1/2} + 12 N_1}{1920 \pi^2} \eqend{,} \\
b' &= - \frac{N_0 + 11 N_{1/2} + 62 N_1}{5760 \pi^2} \eqend{.}
\end{align}
\end{subequations}
(In the literature, also $a = - (4\pi)^2 b'$ and $c = (4\pi)^2 b$ are used, which we do not employ to avoid confusion with the scale factor $a$.) In the conformally flat case, the functional integral in~\eqref{effective_action} can be performed by rescaling the scalar matter as $\tilde{\phi} = a^{1-n/2} \phi$ and working in flat space. Since we are working with conformal matter where the bare action is invariant under a rescaling, we have
\begin{equation}
S_\text{M}\big[ \tilde{g}, \tilde{\phi} \big] = S_\text{M}\!\left[ g, \phi \right]
\end{equation}
and thus also
\begin{equation}
\Sigma\!\left[ \tilde{g}^\pm \right] = \Sigma\!\left[ g^\pm \right] \eqend{,}
\end{equation}
but the counterterms contained in $S_\text{CT}$ are not so simply related. That is, while in the sum
\begin{equation}
\Sigma\!\left[ g^\pm \right] + S_\text{CT}\!\left[ g^+ \right] - S_\text{CT}\!\left[ g^- \right]
\end{equation}
the counterterms cancel the divergences of $\Sigma$ as $n \to 4$ such that the limit is finite, we want to calculate
\begin{equation}
\label{sigma_conformal}
\begin{split}
\Sigma\!\left[ \tilde{g}^\pm \right] + S_\text{CT}\!\left[ \tilde{g}^+ \right] - S_\text{CT}\!\left[ \tilde{g}^- \right] &= \Sigma\!\left[ g^\pm \right] + S_\text{CT}\!\left[ g^+ \right] - S_\text{CT}\!\left[ g^- \right] \\
&\quad+ \left( S_\text{CT}\!\left[ \tilde{g}^+ \right] - S_\text{CT}\!\left[ g^+ \right] \right) - \left( S_\text{CT}\!\left[ \tilde{g}^- \right] - S_\text{CT}\!\left[ g^- \right] \right) \eqend{.}
\end{split} \raisetag{2.2\baselineskip}
\end{equation}
Since also in flat space the counterterms cancel the divergences of $\Sigma$ as $n \to 4$, the terms in the last line do not diverge, but also do not vanish. They therefore give a finite contribution to the effective action, which can be calculated using the conformal transformations given in Appendix~\ref{appendix_conformal}. This is the way how the correct trace anomaly arises in dimensional regularization~\cite{mazurmottola2001}.

For the counterterms in the MS scheme~\eqref{ms_scheme}, we use Eqns.~\eqref{weyl_conformal} and~\eqref{euler_conformal} to obtain
\begin{equation}
\label{counterterms_diff}
\begin{split}
\lim_{n \to 4} \left( S^\text{MS}_\text{CT}[ \tilde{g} ] - S^\text{MS}_\text{CT}[ g ] \right) &= \int \bigg[ \left( b C^2 + b' \mathcal{E}_4 \right) \ln a - 4 b' G^{\mu\nu} a^{-2} ( \nabla_\mu a ) ( \nabla_\nu a ) \\
&\qquad\qquad+ 2 b' a^{-4} \left( ( \nabla a )^2 - 2 a \nabla^2 a \right) ( \nabla a )^2 \bigg] \sqrt{-g} \total^4 x
\end{split}
\end{equation}
with $\nabla^2 = \nabla^\mu \nabla_\mu$, and where we defined the abbreviation
\begin{equation}
( \nabla a )^2 \equiv ( \nabla^\mu a ) ( \nabla_\mu a ) \eqend{.}
\end{equation}
For a single massless, conformally coupled scalar where $b = 1/(1920 \pi^2)$ and $b' = -b/3$ according to Eq.~\eqref{coeffs_b}, this gives
\begin{equation}
\begin{split}
b \tilde{C}^2 + b' \tilde{\mathcal{E}}_4 &= \frac{1}{2880 \pi^2} \left( \tilde{R}^{\alpha\beta\gamma\delta} \tilde{R}_{\alpha\beta\gamma\delta} - \tilde{R}^{\alpha\beta} \tilde{R}_{\alpha\beta} \right) \\
&\quad+ \frac{n-4}{2880 \pi^2} \left( \frac{3}{4} \tilde{C}^2 - \frac{3}{4} \tilde{\mathcal{E}}_4 + \frac{1}{12} \tilde{R}^2 \right) + \bigo{(n-4)^2} \eqend{,}
\end{split}
\end{equation}
such that the scheme used in Refs.~\cite{camposverdaguer1994,camposverdaguer1996} differs by a finite renormalization from the MS scheme:
\begin{equation}
\label{s_ct_ms_cv}
S^\text{MS}_\text{CT} = S^\text{CV}_\text{CT} + \frac{1}{3840 \pi^2} \int \left( \tilde{C}^2 - \tilde{\mathcal{E}}_4 + \frac{1}{9} \tilde{R}^2 \right) \sqrt{-\tilde{g}} \total^4 x \eqend{.}
\end{equation}
While the integral of $\tilde{\mathcal{E}}_4$ is a topological invariant in four dimensions and thus does not contribute to the dynamics, and the finite renormalization of the square of the Weyl tensor can be absorbed simply into a redefinition of the renormalization scale $\bar{\mu}$ in the explicit result for the effective action $S_\text{eff}$ given in Refs.~\cite{camposverdaguer1994,camposverdaguer1996}, the finite renormalization of the square of the Ricci scalar must be taken into account when we generalize the result for $S_\text{eff}$ to general conformal theories.

As explained above, the counterterms cancel the divergences in $\Sigma$ such that $S_\text{eff}$ is finite as $n \to 4$. Including an arbitrary finite coefficient $\beta$ for the square of the Ricci scalar, we obtain the effective action for general conformal theories by comparing formulas~\eqref{seff_in_sloc_sigma},~\eqref{sloc_def},~\eqref{sigma_conformal},~\eqref{counterterms_diff} and~\eqref{s_ct_ms_cv} with the result of Refs.~\cite{camposverdaguer1994,camposverdaguer1996}. The effective action is then given by
\begin{equation}
\label{seff}
S_\text{eff}\!\left[ \tilde{g}^\pm \right] = S_\text{loc,ren}\!\left[ a, g^+ \right] - S_\text{loc,ren}\!\left[ a, g^- \right] + \Sigma_\text{ren}\!\left[ g^\pm \right] \eqend{,}
\end{equation}
where $S_\text{loc,ren}\!\left[ a, g^+ \right]$ is the renormalized, finite part of the local action $S_\text{loc}$~\eqref{sloc_def}, given by
\begin{equation}
\label{slocren_def}
\begin{split}
S_\text{loc,ren}[a, g] &\equiv \frac{1}{\kappa^2} \int \left( a^2 R - 6 a \nabla^2 a - 2 \Lambda a^4 \right) \sqrt{-g} \total^4 x + \int \left( b C^2 + b' \mathcal{E}_4 \right) \ln a \sqrt{-g} \total^4 x \\
&\quad+ \frac{\beta}{12} \int \left( R - 6 a^{-1} \nabla^2 a \right)^2 \sqrt{-g} \total^4 x - 4 b' \int G^{\mu\nu} a^{-2} ( \nabla_\mu a ) ( \nabla_\nu a ) \sqrt{-g} \total^4 x \\
&\quad+ 2 b' \int a^{-4} \left[ ( \nabla a )^2 - 2 a \nabla^2 a \right] ( \nabla a )^2 \sqrt{-g} \total^4 x + S_\text{PP} \eqend{,} \raisetag{1.1\baselineskip}
\end{split}
\end{equation}
and $\Sigma_\text{ren}$ is the renormalized, finite part of the non-local term $\Sigma$, which reads
\begin{equation}
\label{sigmaren_def}
\begin{split}
\Sigma_\text{ren}\!\left[ g^\pm \right] &\equiv 2 b \iint C^+_{\mu\nu\rho\sigma}(x) C^{-\mu\nu\rho\sigma}(y) K(x-y) \total^4 x \total^4 y \\
&\quad+ b \iint C^+_{\mu\nu\rho\sigma}(x) C^{+\mu\nu\rho\sigma}(y) K^+(x-y; \bar{\mu}) \total^4 x \total^4 y \\
&\quad- b \iint C^-_{\mu\nu\rho\sigma}(x) C^{-\mu\nu\rho\sigma}(y) K^-(x-y; \bar{\mu}) \total^4 x \total^4 y
\end{split}
\end{equation}
with the kernels $K$ given by their Fourier transforms
\begin{subequations}
\label{kernels_def}
\begin{align}
K(x) &\equiv - \mathi \pi \int \Theta(-p^2) \Theta(-p^0) \mathe^{\mathi p x} \frac{\total^4 p}{(2\pi)^4} \eqend{,} \\
K^\pm(x; \bar{\mu}) &\equiv \frac{1}{2} \int \left[ - \ln \abs{\frac{p^2}{\bar{\mu}^2}} \pm \mathi \pi \Theta(-p^2) \right] \mathe^{\mathi p x} \frac{\total^4 p}{(2\pi)^4} \eqend{.}
\end{align}
\end{subequations}

This result is valid in the MS renormalization scheme~\eqref{ms_scheme}, where the renormalization scale $\bar{\mu}$ is chosen such that there is no term proportional to $C^2$ in the local part of the renormalized effective action $S_\text{loc,ren}$~\eqref{slocren_def} (except for the term involving $\ln a$ coming from the conformal transformation). However, the effective action is invariant under the renormalization group~\cite{toms1983} and cannot depend on the renormalization scale $\mu$. Thus, for all values of $\mu \neq \bar{\mu}$, an additional term of the form $c \int C^2 \total^4 x$ appears in $S_\text{loc,ren}$, with a coefficient $c$ which depends on the renormalization scale. Noting that the renormalized $S_\text{eff}$ only depends on $\bar{\mu}$ through $\Sigma_\text{ren}$~\eqref{sigmaren_def}, and that the $\bar{\mu}$-dependent part of the kernels $K$ reads
\begin{equation}
\frac{1}{2} \int \left[ - \ln \frac{1}{\bar{\mu}^2} \right] \mathe^{\mathi p x} \frac{\total^4 p}{(2\pi)^4} = \delta^4(x-x') \ln \bar{\mu} \eqend{,}
\end{equation}
the dependence of $S_\text{eff}$ on $\bar{\mu}$ is given by
\begin{equation}
b \ln \bar{\mu} \left[ \int C^+_{\mu\nu\rho\sigma} C^{+\mu\nu\rho\sigma} \total^4 x - \int C^-_{\mu\nu\rho\sigma} C^{-\mu\nu\rho\sigma} \total^4 x \right] \eqend{.}
\end{equation}
Adding the term $c(\mu) \int C^2 \total^4 x$ to $S_\text{loc,ren}$, we then can determine the dependence of $c$ on $\mu$ by demanding that $\total S_\text{eff}\!\left[ \tilde{g}^\pm \right] / \total \mu = 0$, which gives
\begin{equation}
c(\mu) = c(\mu_0) - b \ln \frac{\mu}{\mu_0}
\end{equation}
for some reference scale $\mu_0$ (e.g., the Hubble scale $H$). While we will employ $\bar{\mu}$ in the following to shorten the formulas [since by definition of the MS scheme $c(\bar{\mu}) = 0$], we will restore the finite coefficient $c$ in the final results, i.e., perform the replacement
\begin{equation}
\label{barmu_replace}
b \ln \bar{\mu} \to b \ln \mu + c(\mu) \eqend{.}
\end{equation}

The above result for the renormalized effective action is valid for general conformally flat backgrounds. However, we are interested in cosmological applications, and thus specialize to spatially flat Friedmann-Lema{\^\i}tre-Robertson-Walker (FLRW) backgrounds where the conformal/scale factor $a(\eta)$ appearing in the metric~\eqref{conf_metric} only depends on the conformal time $\eta$, and the metric~\eqref{conf_metric} reads
\begin{equation}
\label{flrw_pert}
\tilde{g}_{\mu\nu} = a^2(\eta) g_{\mu\nu} = a^2(\eta) \left( \eta_{\mu\nu} + h_{\mu\nu} \right) \eqend{.}
\end{equation}
From now on, a prime denotes a derivative with respect to conformal time, and we define the Hubble parameter $H$ by
\begin{equation}
\label{hubble_def}
H \equiv \frac{a'}{a^2} \eqend{.}
\end{equation}

\subsection{Metric perturbations}

The expression~\eqref{seff} for the effective action is valid to second order in metric perturbations around the FLRW background~\eqref{flrw_pert}, and using the expansions from Appendix~\ref{appendix_metric} one can obtain the explicit form in terms of the perturbation. It is convenient to decompose the metric perturbation in irreducible components according to their transformation properties under spatial rotations and translations (with respect to the FLRW background), given by
\begin{subequations}
\label{irreducible}
\begin{align}
h_{00} &= 2 s_1 \eqend{,} \\
h_{0k} &= v^{\text{T}1}_k + \partial_k s_2 \eqend{,} \\
h_{kl} &= h^\text{TT}_{kl} + 2 \partial_{(k} v^{\text{T}2}_{l)} + 2 \left( \partial_k \partial_l - \frac{\delta_{kl} \laplace}{3} \right) s_3 + 2 \delta_{kl} s_4 \eqend{,}
\end{align}
\end{subequations}
where $\laplace = \partial_k \partial_k$, the two vectors $v^{\text{T}i}_k$ are transverse, $\partial_k v^{\text{T}i}_k = 0$, and the tensor $h^\text{TT}_{kl}$ is transverse and traceless, $\partial_k h^\text{TT}_{kl} = h^\text{TT}_{kk} = 0$ (since the spatial metric is the identity, we do not make a distinction between lower and upper indices). Under an infinitesimal coordinate transformation $x^\mu \to x^\mu + \delta x^\mu = x^\mu + \xi^\mu$, we have
\begin{equation}
\delta h_{\mu\nu} = a^{-2} \mathcal{L}_{a^2 \xi} \left( a^2 \eta_{\mu\nu} \right) = 2 \partial_{(\mu} \xi_{\nu)} - 2 H a \eta_{\mu\nu} \xi_0 \eqend{,}
\end{equation}
where the indices of $\xi^\mu$ are raised and lowered with the flat metric. For the irreducible components~\eqref{irreducible}, this change reads
\begin{subequations}
\begin{align}
\delta s_1 &= \xi_0' + H a \xi_0 \eqend{,} \quad \delta s_2 = \xi_0 + \frac{\partial_k}{\laplace} \xi_k' \eqend{,} \\
\delta s_3 &= \frac{\partial_k}{\laplace} \xi_k \eqend{,} \quad \delta s_4 = \frac{1}{(n-1)} \partial_k \xi_k - H a \xi_0 \eqend{,} \\
\delta v^\text{T1}_k &= \xi_k' - \frac{\partial_k \partial_l}{\laplace} \xi_l' \eqend{,} \quad \delta v^\text{T2}_k = \xi_k - \frac{\partial_k \partial_l}{\laplace} \xi_l \eqend{,} \\
\delta h^\text{TT}_{kl} &= 0 \eqend{.}
\end{align}
\end{subequations}
Note that here we only consider perturbations and gauge transformations of finite extent (or sufficiently rapid decay at spatial infinity), such that the inverse of the Laplace operator (with vanishing boundary conditions) is well defined. We see that from these components we can form two gauge-invariant scalars, the Bardeen potentials~\cite{bardeen1980}
\begin{subequations}
\begin{align}
\Phi_\text{A} &\equiv s_1 - ( s_2 - s_3' )' - H a ( s_2 - s_3' ) \eqend{,} \\
\Phi_\text{H} &\equiv s_4 - \frac{1}{3} \laplace s_3 + H a ( s_2 - s_3' )
\end{align}
\end{subequations}
and one gauge-invariant vector
\begin{equation}
V_k \equiv v^\text{T1}_k - v^{\text{T2}\prime}_k \eqend{,} \qquad V_0 \equiv 0 \eqend{,}
\end{equation}
while the tensor part $h^\text{TT}_{kl}$ is automatically gauge invariant. The remaining components can be arranged into a vector $X^\mu$ defined by
\begin{subequations}
\begin{align}
X^0 &= s_3' - s_2 \eqend{,} \\
X^k &= \eta^{kl} \left( v^\text{T2}_l + \partial_l s_3 \right) \eqend{,}
\end{align}
\end{subequations}
which has the simple gauge transformation
\begin{equation}
\delta X^\mu = \xi^\mu \eqend{.}
\end{equation}
Thus the metric perturbation can be written in the form~\cite{abramobrandenbergermukhanov1997,nakamura2007}
\begin{equation}
\begin{split}
\label{h_munu_inv_gauge}
h_{\mu\nu} &= h^\text{inv}_{\mu\nu} + a^{-2} \mathcal{L}_{a^2 X} \left( a^2 \eta_{\mu\nu} \right) \\
&= h^\text{inv}_{\mu\nu} + 2 \partial_{(\mu} X_{\nu)} - 2 H a \eta_{\mu\nu} X_0 \eqend{,}
\end{split}
\end{equation}
where the gauge-invariant part
\begin{equation}
h^\text{inv}_{\mu\nu} \equiv 2 \delta^0_\mu \delta^0_\nu \left( \Phi_\text{A} + \Phi_\text{H} \right) + 2 \eta_{\mu\nu} \Phi_\text{H} + 2 \delta^0_{(\mu} V_{\nu)} + h^\text{TT}_{\mu\nu}
\end{equation}
with $h^\text{TT}_{0\mu} \equiv 0$ does not change under infinitesimal coordinate transformations.

\subsection{The in-in effective action}

We now insert the decomposition~\eqref{h_munu_inv_gauge} into the effective action $S_\text{eff}$. Since we only want to calculate the corrections to the Newtonian potential due to quantum effects, we neglect the backreaction of the point particle on the metric, and consequently only expand the point particle action $S_\text{PP}$ of Eq.~\eqref{action_pp} to first order in perturbations. For the same reason, the background equations of motion -- the semiclassical Einstein equations -- must not take into account the presence of the point particle. These equations are obtained by taking a variational derivative of the effective action with respect to the ``$+$'' perturbation and setting it to zero afterwards, which for the FLRW background~\eqref{flrw_pert} reads
\begin{equation}
\label{bg_eom_h}
\left. \frac{\delta S_\text{eff}[ a, h^\pm ]}{\delta h_{\mu\nu}^+} \right\rvert_{h^\pm = 0} = 0 \eqend{.}
\end{equation}
Since the quantum state of the matter fields which appear in the functional integral~\eqref{seff_funcint} depends on the background metric, one must find a self-consistent solution to these equations, which in the general case may be very hard. In our case, the conformal matter fields are in the conformal vacuum state, and the semiclassical Einstein equation~\eqref{bg_eom_h} determines the scale factor $a(\eta)$. For the effective action~\eqref{seff}, ignoring $S_\text{PP}$ as explained above, we obtain
\begin{equation}
a(\eta) = - \frac{1}{H\eta}
\end{equation}
with the Hubble parameter $H$ defined in Eq.~\eqref{hubble_def} determined in terms of the cosmological constant $\Lambda$ (and vice versa). Generalizing the result from Ref.~\cite{frv2012} to a general conformal theory, this is
\begin{equation}
\label{lambda_h}
\Lambda = 3 H^2 \left( 1 + b' \kappa^2 H^2 \right) \eqend{,}
\end{equation}
with constant $H$. Thus, the quantum-corrected background is a de~Sitter spacetime, albeit with a slightly different radius as compared to the classical one.

Expanding now the point particle action to first order, we obtain
\begin{equation}
\label{s_pp_expand}
S_\text{PP} = S_\text{PP}[a, g] + \frac{1}{2} \int a^2 h_{\mu\nu} T^{\mu\nu} \sqrt{-g} \total^4 x + \bigo{h^2} \eqend{,}
\end{equation}
where
\begin{equation}
T^{\mu\nu} = \frac{m}{\sqrt{-g}} \int \frac{\total z^\mu}{\total \tau} \frac{\total z^\nu}{\total \tau} \delta^4(x-z(\tau)) \total \tau
\end{equation}
is the stress tensor of the point particle. Considering a particle at rest at the origin, we have
\begin{equation}
\frac{\total z^\mu}{\total \tau} = \delta^\mu_0 a^{-1} \eqend{,}
\end{equation}
which is normalized with the background metric
\begin{equation}
a^2 \eta_{\mu\nu} \frac{\total z^\mu}{\total \tau} \frac{\total z^\nu}{\total \tau} = - 1 \eqend{,}
\end{equation}
and thus
\begin{equation}
T^{\mu\nu} = m a^{-6}(\eta) \delta^\mu_0 \delta^\nu_0 \int \delta(\eta-z^0(\tau)) \total \tau \delta^3(\vec{x}) \eqend{.}
\end{equation}
In this case, we have
\begin{equation}
\total \tau^2 = a^2(\eta) \total \eta^2 \eqend{,}
\end{equation}
such that
\begin{equation}
\int \delta(\eta-z^0(\tau)) \total \tau = \int a(\eta) \delta(\eta-z^0(\tau)) \total \eta = a(\eta)
\end{equation}
and (see also Ref.~\cite{iliopoulosetal1998})
\begin{equation}
\label{t_munu}
T^{\mu\nu}(\eta,\vec{x}) = m a^{-5} \delta^\mu_0 \delta^\nu_0 \delta^3(\vec{x}) \eqend{.}
\end{equation}
It can be easily checked that this stress tensor satisfies the correct covariant conservation law $\nabla^{(0)}_\mu T^{\mu\nu} = 0$, where the covariant derivative $\nabla^{(0)}_\mu$ is the one associated to the background metric $g^{(0)}_{\mu\nu} \equiv a^2 \eta_{\mu\nu}$. Inserting the explicit expression~\eqref{t_munu} back into the expansion~\eqref{s_pp_expand} and using the decomposition~\eqref{h_munu_inv_gauge}, the linear part reads
\begin{equation}
\label{seff_pp}
\frac{m}{2} \int a h_{00} \delta^3(\vec{x}) \total^4 x = m \int a \Phi_\text{A} \delta^3(\vec{x}) \total^4 x + m \int a ( X_0' + H a X_0 ) \delta^3(\vec{x}) \total^4 x \eqend{.}
\end{equation}
Recalling the definition of the Hubble parameter~\eqref{hubble_def}, the second term can be written as a total derivative
\begin{equation}
a ( X_0' + H a X_0 ) = ( a X_0 )' \eqend{,}
\end{equation}
and does not contribute to the effective action. To the lowest order we are working, the coupling is thus gauge-invariant, and the results we will obtain have a direct physical meaning. Moreover, we see that the point particle at rest only couples to scalar perturbations. We can therefore ignore the vector and tensor perturbations, since their equations of motion will not change from the source-free case treated in Ref.~\cite{fprv2013} and they do not contribute to the gravitational potential of our point particle. For the scalar part of $S_\text{loc,ren}$ [referring to the scalar part with a superscript $(\text{S})$], we then obtain after a long but straightforward calculation using the relation~\eqref{lambda_h}, the expansions from Appendix~\ref{appendix_metric} and the decomposition~\eqref{h_munu_inv_gauge} the following expression:
\begin{equation}
\label{seff_bardeen}
\begin{split}
S^{(\text{S})}_\text{loc,ren} &= m \int a \Phi_\text{A} \delta^3(\vec{x}) \total^4 x - \frac{2}{\kappa^2} \int a^2 \left( 3 + 10 b' \kappa^2 H^2 + 6 \beta \kappa^2 H^2 \right) \left( \Phi_\text{H}' + H a \Phi_\text{A} \right)^2 \total^4 x \\
&\quad- \frac{2}{\kappa^2} \int a^2 \left( 1 + 6 b' \kappa^2 H^2 + 2 \beta \kappa^2 H^2 \right) \Phi_\text{H} \laplace \left( \Phi_\text{H} - 2 \Phi_\text{A} \right) \total^4 x \\
&\quad+ \frac{4 b}{3} \int \laplace \left( \Phi_\text{A} + \Phi_\text{H} \right) \laplace \left( \Phi_\text{A} + \Phi_\text{H} \right) \ln a \total^4 x \\
&\quad+ \frac{\beta}{3} \int \Big[ 3 \left( \Phi_\text{H}' + H a \Phi_\text{A} \right)' + 9 H a \left( \Phi_\text{H}' + H a \Phi_\text{A} \right) + \laplace \left( \Phi_\text{A} - 2 \Phi_\text{H} \right) \Big]^2 \total^4 x \eqend{.} \raisetag{1.2\baselineskip}
\end{split}
\end{equation}
Furthermore, for the renormalized non-local part $\Sigma_\text{ren}$~\eqref{sigmaren_def} we obtain the expansion
\begin{equation}
\label{seff_bardeen2}
\begin{split}
\Sigma^{(\text{S})}_\text{ren} &= \frac{8 b}{3} \iint \left[ \laplace \Phi_\text{A}^+(x) + \laplace \Phi_\text{H}^+(x) \right] K(x-y) \left[ \laplace \Phi_\text{A}^-(y) + \laplace \Phi_\text{H}^-(y) \right] \total^4 x \total^4 y \\
&\quad+ \frac{4 b}{3} \iint \left[ \laplace \Phi_\text{A}^+(x) + \laplace \Phi_\text{H}^+(x) \right] K^+(x-y; \bar{\mu}) \left[ \laplace \Phi_\text{A}^+(y) + \laplace \Phi_\text{H}^+(y) \right] \total^4 x \total^4 y \\
&\quad- \frac{4 b}{3} \iint \left[ \laplace \Phi_\text{A}^-(x) + \laplace \Phi_\text{H}^-(x) \right] K^-(x-y; \bar{\mu}) \left[ \laplace \Phi_\text{A}^-(y) + \laplace \Phi_\text{H}^-(y) \right] \total^4 x \total^4 y \eqend{.}
\end{split}
\end{equation}
This result can be checked by comparing with the expression given in Ref.~\cite{frv2014}. There, $N$ free, conformally coupled scalar fields were treated such that $b = 3/2 \alpha$ and $b' = - \alpha/2$ with $\alpha = N/(2880 \pi^2)$ (although later on the generalization of the results, but not the action, to general conformal theories was made), and the gauge choice taken in this work corresponds to making a gauge transformation such that
\begin{equation}
X_k = 0 \eqend{,} \qquad X_0 = (Ha)^{-1} \Phi_\text{H} \eqend{.}
\end{equation}
The relation between the Bardeen potentials and the scalars $\phi$ and $\psi$ employed by Ref.~\cite{frv2014} is then given by
\begin{equation}
\Phi_\text{H} = \kappa H a \psi \eqend{,} \quad \Phi_\text{A} = \kappa \left( \frac{1}{2} \phi - \psi' - H a \psi \right) \eqend{.}
\end{equation}

\section{Field equations}
\label{sec_fieldequations}

The effective field equations are obtained by taking a functional derivative of the effective action $S_\text{eff}[ a, h^\pm ]$ with respect to the ``$+$'' fields and setting both types of fields to be equal,
\begin{equation}
\left. \frac{\delta S_\text{eff}[ a, h^\pm ]}{\delta h^+} \right\rvert_{h^+ = h^- = h} = 0 \eqend{.}
\end{equation}
As explained above, since the particle at rest only couples to scalar perturbations, we ignore the vector and tensor parts of the effective action. Relatively simple equations are then obtained by taking $\delta S^{(\text{S})}_\text{eff}/\delta \Phi_\text{A} = 0$, which gives
\begin{equation}
\label{semiclassical_eqna}
\laplace \Phi_\text{H} - 3 H a \left( \Phi_\text{H}' + H a \Phi_\text{A} \right) - \kappa^2 H^2 \mathcal{S}_1[ \Phi_\text{A}, \Phi_\text{H} ] = - \frac{\kappa^2}{4 a} m \delta^3(\vec{x})
\end{equation}
with
\begin{equation}
\label{source_1}
\begin{split}
\mathcal{S}_1[ \Phi_\text{A}, \Phi_\text{H} ] &= - 6 b' \laplace \Phi_\text{H} + 10 b' H a \left( \Phi_\text{H}' + H a \Phi_\text{A} \right) \\
&\quad+ \frac{1}{2} \beta (Ha)^{-2} \left[ \left( \partial_\eta + 3 H a \right) \laplace - 3 H a \left( \partial_\eta^2 - 2 H^2 a^2 \right) \right] \left( \Phi_\text{H}' + H a \Phi_\text{A} \right) \\
&\quad+ \frac{1}{6} \beta (Ha)^{-2} \laplace \left( 3 H a \partial_\eta - 3 H^2 a^2 - \laplace \right) \left( 2 \Phi_\text{H} - \Phi_\text{A} \right) + \beta \laplace \Phi_\text{A} \\
&\quad- \frac{2}{3} b (Ha)^{-2} \laplace \int \laplace \left( \Phi_\text{A} + \Phi_\text{H} \right)(x') \left( H(x-x'; \bar{\mu}) + \delta^4(x-x') \ln a \right) \total^4 x' \eqend{,}
\end{split}
\end{equation}
and $\delta S^{(\text{S})}_\text{eff}/\delta \Phi_\text{A} + \delta S^{(\text{S})}_\text{eff}/\delta \Phi_\text{H} = 0$, which gives
\begin{equation}
\label{semiclassical_eqnb}
\laplace \Phi_\text{A} + 3 \left( \Phi_\text{H}' + H a \Phi_\text{A} \right)' + 3 H a \left( \Phi_\text{H}' + H a \Phi_\text{A} \right) - \kappa^2 H^2 \mathcal{S}_2[ \Phi_\text{A}, \Phi_\text{H} ] = - \frac{\kappa^2}{4 a} m \delta^3(\vec{x}) \\
\end{equation}
with
\begin{equation}
\label{source_2}
\begin{split}
\mathcal{S}_2[ \Phi_\text{A}, \Phi_\text{H} ] &= - 6 b' \laplace \Phi_\text{A} - 10 b' \left( \partial_\eta + H a \right) \left( \Phi_\text{H}' + H a \Phi_\text{A} \right) - 12 \beta H a \left( \Phi_\text{H}' + H a \Phi_\text{A} \right) \\
&\quad- \frac{1}{2} \beta (Ha)^{-2} \left( \laplace - 3 \partial_\eta^2 + 12 H a \partial_\eta - 12 H^2 a^2 \right) \left( \partial_\eta + 3 H a \right) \left( \Phi_\text{H}' + H a \Phi_\text{A} \right) \\
&\quad+ \frac{1}{6} \beta (Ha)^{-2} \laplace \left( - 81 \partial_\eta^2 + 12 H a \partial_\eta + \laplace \right) \left( 2 \Phi_\text{H} - \Phi_\text{A} \right) + 2 \beta \laplace \Phi_\text{A} \\
&\quad- \frac{4}{3} b (Ha)^{-2} \laplace \int \laplace \left( \Phi_\text{A} + \Phi_\text{H} \right)(x') \left( H(x-x'; \bar{\mu}) + \delta^4(x-x') \ln a \right) \total^4 x' \eqend{.}
\end{split}
\end{equation}
Here, the kernel $H$ is a particular combination of the kernels $K$ defined in Eq.~\eqref{kernels_def}, given by
\begin{equation}
H(x-x'; \bar{\mu}) \equiv K(x-y) + K^+(x-y; \bar{\mu}) \eqend{.}
\end{equation}
Later on it will be convenient to have its spatial Fourier transform
\begin{equation}
\tilde{H}(\eta-\eta', \vec{p}; \bar{\mu}) \equiv \int H(x-x'; \bar{\mu}) \, \mathe^{- \mathi \vec{p} (\vec{x}-\vec{x}')} \total^3 x \eqend{,}
\end{equation}
which has been calculated in Ref.~\cite{frv2012}. It reads
\begin{equation}
\label{kernel_h_fourier}
\tilde{H}(\eta-\eta', \vec{p}; \bar{\mu}) = \cos\left[\abs{\vec{p}} (\eta-\eta')\right] \distlim_{\epsilon \to 0} \left[ \frac{\Theta(\eta-\eta'-\epsilon)}{\eta-\eta'} + \delta(\eta-\eta') \left( \ln(\bar{\mu} \epsilon) + \gamma \right) \right] \eqend{,}
\end{equation}
where $\gamma$ is the Euler-Mascheroni constant, and the notation ``$\distlim$'' means that the limit $\epsilon \to 0$ is to be taken in the sense of distributions, i.e., after integration with a test function.

While these equations involve only the two Bardeen potentials, each of those has two different contributions. The first one is given by the classical response of the gravitational field to a test particle, which in the flat-space limit reduces to the classical Newtonian potential, and which comes from the Einstein-Hilbert gravitational action $S_\text{G}$ of Eq.~\eqref{action_eh}. The second contribution is the one we are really interested in, which incorporates the quantum effects due to the vacuum polarization of conformal matter, and which is sourced by the classical contribution. This contribution is suppressed by a factor of $\kappa^2$, as can be seen from the explicit form~\eqref{seff_bardeen} of the effective action, where the quantum contributions (the terms which involve $b$, $b'$ or $\beta$) are of order $\kappa^0$, while the Einstein-Hilbert term is of order $\kappa^{-2}$. We thus split the Bardeen potentials into a classical and a quantum contribution
\begin{equation}
\label{bardeen_split}
\Phi_{\text{A}/\text{H}} = \Phi^\text{cl}_{\text{A}/\text{H}} + \kappa^2 \Phi^\text{qu}_{\text{A}/\text{H}} \eqend{,}
\end{equation}
and from~\eqref{semiclassical_eqna} and~\eqref{semiclassical_eqnb} obtain the equations
\begin{subequations}
\label{eqn_cl}
\begin{align}
\laplace \Phi^\text{cl}_\text{H} - 3 H a \left( \Phi^{\text{cl}\prime}_\text{H} + H a \Phi^\text{cl}_\text{A} \right) &= - \frac{\kappa^2}{4 a} m \delta^3(\vec{x}) \eqend{,} \label{eqn_cl_a} \\
\laplace \Phi^\text{cl}_\text{A} + 3 \left( \Phi^{\text{cl}\prime}_\text{H} + H a \Phi^\text{cl}_\text{A} \right)' + 3 H a \left( \Phi^{\text{cl}\prime}_\text{H} + H a \Phi^\text{cl}_\text{A} \right) &= - \frac{\kappa^2}{4 a} m \delta^3(\vec{x}) \label{eqn_cl_b}
\end{align}
\end{subequations}
for the classical and
\begin{subequations}
\label{eqn_qu}
\begin{align}
\laplace \Phi^\text{qu}_\text{H} - 3 H a \left( \Phi^{\text{qu}\prime}_\text{H} + H a \Phi^\text{qu}_\text{A} \right) &= H^2 \mathcal{S}_1\!\left[ \Phi^\text{cl}_\text{A}, \Phi^\text{cl}_\text{H} \right] \\
\laplace \Phi^\text{qu}_\text{A} + 3 \left( \Phi^{\text{qu}\prime}_\text{H} + H a \Phi^\text{qu}_\text{A} \right)' + 3 H a \left( \Phi^{\text{qu}\prime}_\text{H} + H a \Phi^\text{qu}_\text{A} \right) &= H^2 \mathcal{S}_2\!\left[ \Phi^\text{cl}_\text{A}, \Phi^\text{cl}_\text{H} \right]
\end{align}
\end{subequations}
for the quantum contribution. On the right-hand side of Eqns.~\eqref{eqn_qu}, the source terms $\mathcal{S}_1$ and $\mathcal{S}_2$ in principle should involve the full Bardeen potentials~\eqref{bardeen_split}, including the quantum corrections. However, these terms are of order $\kappa^4$, and do not contribute to the order we are working.

Let us quickly make a comment about the flat-space limit: Taking $a = 1$ and $H = 0$ in Eqns.~\eqref{eqn_cl}, and using that $\kappa^2 = 16 \pi G_\text{N}$, the Bardeen potential $\Phi^\text{cl}_\text{H}$ satisfies the standard Poisson equation. Since its solution is time-independent, the second equation then also reduces to the same Poisson equation for $\Phi^\text{cl}_\text{A}$, which then in the Newtonian limit gives the correct Newtonian potential.

\section{Solutions}
\label{sec_solutions}

Since the equations~\eqref{eqn_cl} and~\eqref{eqn_qu} are coupled, we first have to find a linear combination which allows to solve for one of the Bardeen potentials. Taking a time derivative $\partial_\eta$ of the left-hand side of Eq.~\eqref{eqn_cl_a} and adding $H a$ times the left-hand side of Eq.~\eqref{eqn_cl_b} one obtains
\begin{equation}
\label{eqn_cl_c}
\laplace \left( \Phi^{\text{cl}\prime}_\text{H} + H a \Phi^\text{cl}_\text{A} \right) = 0 \eqend{,}
\end{equation}
and the analogue combination for the quantum contribution reads
\begin{equation}
\label{eqn_qu_c}
\laplace \left( \Phi^{\text{qu}\prime}_\text{H} + H a \Phi^\text{qu}_\text{A} \right) = H^2 \partial_\eta \, \mathcal{S}_1\!\left[ \Phi^\text{cl}_\text{A}, \Phi^\text{cl}_\text{H} \right] + H^3 a \, \mathcal{S}_2\!\left[ \Phi^\text{cl}_\text{A}, \Phi^\text{cl}_\text{H} \right] \eqend{.}
\end{equation}
Once these are solved, we can substitute the solutions back into Eqns.~\eqref{eqn_cl} and~\eqref{eqn_qu} and solve for the individual Bardeen potentials.

\subsection{The classical contribution}

As said before, we consider only metric perturbations which decay sufficiently rapidly at spatial infinity, such that the inverse of the Laplace operator (with vanishing boundary conditions) is well defined. The solution of equation~\eqref{eqn_cl_c} is then given by
\begin{equation}
\label{sol_cl_c}
\Phi^{\text{cl}\prime}_\text{H} + H a \Phi^\text{cl}_\text{A} = 0 \eqend{,}
\end{equation}
and substituting into Eqns.~\eqref{eqn_cl} we obtain
\begin{equation}
\laplace \Phi^\text{cl}_\text{H} = \laplace \Phi^\text{cl}_\text{A} = - \frac{\kappa^2}{4 a} m \delta^3(\vec{x}) \eqend{.}
\end{equation}
Inverting the Laplacian and using the well-known formula
\begin{equation}
\laplace \frac{1}{r} = - 4 \pi \delta^3(\vec{x})
\end{equation}
with $r = \abs{\vec{x}}$, we obtain the solution
\begin{equation}
\label{sol_cl_ab}
\Phi^\text{cl}_\text{H} = \Phi^\text{cl}_\text{A} = \frac{\kappa^2 m}{16 \pi a r} = \frac{G_\text{N} m}{a r} \eqend{.}
\end{equation}
This solution also fulfils equation~\eqref{sol_cl_c}, as can be easily checked. Since $a r$ is the physical distance on equal-time hypersurfaces, the result is time-independent for a fixed distance. We thus conclude that classical systems bound by their own gravitational attraction do not ``feel'' the expansion of the universe (at least in the test particle approximation), a result which has been known for a long time~\cite{einsteinstraus1945}. However, recent studies show that this is only true for an exact de~Sitter background~\cite{adkinsmcdonnellfell2007,faraonijaques2007,carreragiulini2008}, and Wang and Woodard~\cite{wangwoodard2015} and Park, Prokopec and Woodard~\cite{parkprokopecwoodard2015} find a time dependence of the gravitational potentials due to quantum corrections even in exact de~Sitter space. It thus remains to check whether the quantum corrections have an explicit dependence on the scale factor $a$ other than the combination $a r$.

\subsection{The quantum contribution}
\label{sec_solutions_quantum}

Before we can solve Eqns.~\eqref{eqn_qu_c} and~\eqref{eqn_qu}, we must first calculate the source terms $\mathcal{S}_1$ and $\mathcal{S}_2$ defined in Eqns.~\eqref{source_1} and~\eqref{source_2}, which depend on the classical solution~\eqref{sol_cl_ab}. Using that $\Phi^\text{cl}_\text{H} = \Phi^\text{cl}_\text{A}$ [according to Eq.~\eqref{sol_cl_ab}] to replace all occurences of $\Phi^\text{cl}_\text{A}$ by $\Phi^\text{cl}_\text{H}$, and simplifying terms using the relation~\eqref{sol_cl_c}, it is not even necessary to substitute the concrete solution, and we obtain
\begin{equation}
\mathcal{S}_1\!\left[ \Phi^\text{cl}_\text{A}, \Phi^\text{cl}_\text{H} \right] = - 6 b' \laplace \Phi^\text{cl}_\text{H} - \frac{\beta + 8 b \ln a}{6 (Ha)^2} \laplace^2 \Phi^\text{cl}_\text{H} - \frac{4 b}{3 (Ha)^2} \laplace \int \laplace \Phi^\text{cl}_\text{H}(x') H(x-x'; \bar{\mu}) \total^4 x'
\end{equation}
and
\begin{equation}
\mathcal{S}_2\!\left[ \Phi^\text{cl}_\text{A}, \Phi^\text{cl}_\text{H} \right] = - 6 b' \laplace \Phi^\text{cl}_\text{H} + \frac{\beta - 16 b \ln a}{6 (Ha)^2} \laplace^2 \Phi^\text{cl}_\text{H} - \frac{8 b}{3 (Ha)^2} \laplace \int \laplace \Phi^\text{cl}_\text{H}(x') H(x-x'; \bar{\mu}) \total^4 x' \eqend{.}
\end{equation}
To completely determine the source terms, we need to calculate the non-local term
\begin{equation}
\label{nonlocal_int}
\begin{split}
\int \laplace \Phi^\text{cl}_\text{H}(x') H(x-x'; \bar{\mu}) \total^4 x' &= - \frac{\kappa^2}{4} m \int \delta^3(\vec{x}') a^{-1}(x') H(x-x'; \bar{\mu}) \total^4 x' \\
&\equiv - \frac{\kappa^2}{4 a} m \, I(x; \bar{\mu}) \eqend{.}
\end{split}
\end{equation}
This term depends on the entire past history and on the initial state of the coupled quantum system consisting of matter fields and metric perturbations (here in the form of the Bardeen potentials). However, it does \emph{not} depend on the future history, as it would for the usual in-out formalism. This can be seen from the explicit form of the kernel $H$ in Eq.~\eqref{kernel_h_fourier} that vanishes whenever $\eta' > \eta$. We thus see the in-in formalism explicitly at work here, which guarantees causal evolution equations~\cite{schwinger1961,keldysh1964,chousuhaoyu1985}. It remains to specify the initial state.

In general, the vacuum $\ket{0}$ of the free theory (the Bunch-Davies vacuum in de~Sitter spacetime) is modified by interactions. In our case, the perturbation $h_{\mu\nu}$ is really the expectation value $\bra{\Omega} \hat{h}_{\mu\nu} \ket{\Omega}$ of the field operator $\hat{h}_{\mu\nu}$ in the interacting vacuum state $\Omega$ of the coupled quantum system~\cite{jordan1986,jordan1987}. Starting with an initial state at finite time $\eta_0$, and assuming that the vacuum $\ket{0}$ of the free theory has at least some overlap with the vacuum $\ket{\Omega}$ of the interacting theory (since otherwise perturbation theory would be meaningless), the corrections to $\ket{0}$ can be calculated in perturbation theory, and need to be included in the functional integral~\eqref{effective_action} for the effective action (see Refs.~\cite{collinsholman2005,collins2013,collinsholmanverdanyan2014} for a discussion in the cosmological setting). In Minkowski space, these corrections are also present, but there exists a simple prescription to generate them automatically: by tilting the time integration slightly into the imaginary direction, the interaction is adiabatically switched off for early times $\eta_0 \to -\infty$. Starting with the free vacuum at the infinite past but evolving with the full interacting Hamiltonian, one selects in this way a fully interacting ground state as the state of lowest energy of the full Hamiltonian, i.e., the interacting vacuum~\cite{peskinschroeder}. For a time-dependent Hamiltonian, which does not have a ground state, this approach only works under certain conditions (namely, when the behaviour of the free-field modes is dominated at early times by the same oscillatory behaviour as in Minkowski space), where it selects an adiabatic vacuum state. In the Poincar{\'e} patch of de~Sitter space these conditions are fulfilled, and the proper $\mathi \epsilon$ prescription to select the adiabatic interacting vacuum at early times consists in the replacement
\begin{equation}
\eta \to \eta (1 \mp \mathi \epsilon)
\end{equation}
for the ``$+$'' resp. ``$-$'' fields. One can then take the limit $\eta_0 \to - \infty$ after performing the time integrations and obtains a finite result. Since the kernel $H(x-x'; \bar{\mu})$ appearing in the non-local term is a mixture of loop corrections from ``$+$'' and ``$-$'' fields, this $\mathi \epsilon$ prescription is not straightforward to implement. By carefully following the computations of Refs.~\cite{camposverdaguer1994,camposverdaguer1996} and the spatial Fourier transforms calculated in Ref.~\cite{frv2012}, one sees that the effect of the $\mathi \epsilon$ prescription is to multiply the spatial Fourier transform of the kernel $\tilde{H}(\eta-\eta', \vec{p}; \bar{\mu})$ by a factor $\exp\left[ - \mathi \epsilon \abs{\vec{p}} (\eta-\eta') \right]$.

For this choice of the initial state, the integral $I(x; \bar{\mu})$ is calculated in Appendix~\ref{appendix_nonlocal}, and only depends on $r$ but not on conformal time $\eta$, such that the sources reduce to
\begin{equation}
\mathcal{S}_1\!\left[ \Phi^\text{cl}_\text{A}, \Phi^\text{cl}_\text{H} \right] = - 6 b' \laplace \Phi^\text{cl}_\text{H} - \frac{\beta + 8 b \ln a}{6 (Ha)^2} \laplace^2 \Phi^\text{cl}_\text{H} + \frac{b}{3 (Ha)^2} \frac{\kappa^2}{a} m \laplace I(r; \bar{\mu})
\end{equation}
and
\begin{equation}
\mathcal{S}_2\!\left[ \Phi^\text{cl}_\text{A}, \Phi^\text{cl}_\text{H} \right] = - 6 b' \laplace \Phi^\text{cl}_\text{H} + \frac{\beta - 16 b \ln a}{6 (Ha)^2} \laplace^2 \Phi^\text{cl}_\text{H} + \frac{2 b}{3 (Ha)^2} \frac{\kappa^2}{a} m \laplace I(r; \bar{\mu}) \eqend{.}
\end{equation}
Using again the relation~\eqref{sol_cl_c}, equation~\eqref{eqn_qu_c} then reads
\begin{equation}
\laplace \left( \Phi^{\text{qu}\prime}_\text{H} + H a \Phi^\text{qu}_\text{A} \right) = \frac{2}{3} \left( \beta - 2 b + 2 b \ln a \right) H a^{-1} \laplace^2 \Phi^\text{cl}_\text{H} - \frac{1}{3} b H a^{-2} \kappa^2 m \laplace I(r; \bar{\mu}) \eqend{.}
\end{equation}
Applying the inverse Laplace operator to obtain an expression for $\Phi^{\text{qu}\prime}_\text{H} + H a \Phi^\text{qu}_\text{A}$, we can substitute it into Eqns.~\eqref{eqn_qu} to obtain
\begin{equation}
\label{eqn_qu_a2}
\begin{split}
\laplace \Phi^\text{qu}_\text{H} &= 2 H^2 \left( \beta - 2 b - 3 b' + 2 b \ln a \right) \laplace \Phi^\text{cl}_\text{H} - \frac{1}{6} \left( \beta + 8 b \ln a \right) a^{-2} \laplace^2 \Phi^\text{cl}_\text{H} \\
&\quad- b H^2 a^{-1} \kappa^2 m I(r; \bar{\mu}) + \frac{1}{3} b a^{-3} \kappa^2 m \laplace I(r; \bar{\mu})
\end{split}
\end{equation}
and
\begin{equation}
\label{eqn_qu_b2}
\begin{split}
\laplace \Phi^\text{qu}_\text{A} &= 2 H^2 \left( \beta - 4 b - 3 b' + 2 b \ln a \right) \laplace \Phi^\text{cl}_\text{H} + \frac{1}{6} \left( \beta - 16 b \ln a \right) a^{-2} \laplace^2 \Phi^\text{cl}_\text{H} \\
&\quad- b H^2 a^{-1} \kappa^2 m I(r; \bar{\mu}) + \frac{2}{3} b a^{-3} \kappa^2 m \laplace I(r; \bar{\mu}) \eqend{.}
\end{split}
\end{equation}
Again, to obtain explicit expressions one only has to apply the inverse Laplace operator, and substitute the result~\eqref{sol_cl_ab} for $\Phi^\text{cl}_\text{H}$ and~\eqref{appendix_nonlocal_i} for $I(r; \bar{\mu})$ from the Appendix, which reads
\begin{equation}
I(r; \bar{\mu}) = - \laplace \frac{\ln \left( \mathe^\gamma \bar{\mu} r \right)}{4 \pi r} \eqend{.}
\end{equation}
The result is a well-defined distribution in three spatial dimensions, including the origin $r = 0$. However, for very small $r$ the test particle approximation breaks down (since there the particle's own gravitational field is strong and we cannot neglect the backreaction anymore), and we thus restrict to $r > 0$. There are some local terms supported at the origin [e.g., the terms involving $\laplace^2 \Phi^\text{cl}_\text{H}$ in Eqns.~\eqref{eqn_qu_a2} and~\eqref{eqn_qu_b2}] which then do not contribute. Furthermore, we can use
\begin{equation}
\laplace \frac{\ln r}{r} = - \frac{1}{r^3} \qquad (r>0) \eqend{,}
\end{equation}
and it follows that
\begin{equation}
\label{sol_qu_a}
\Phi^\text{qu}_\text{H} = H^2 \left( \beta - 2 b - 3 b' \right) \frac{\kappa^2 m}{8 \pi a r} + b H^2 \kappa^2 m \frac{\ln \left( \mathe^\gamma \bar{\mu} a r \right)}{4 \pi a r} + \frac{b}{3} \frac{\kappa^2 m}{4 \pi a^3 r^3}
\end{equation}
and
\begin{equation}
\label{sol_qu_b}
\Phi^\text{qu}_\text{A} = H^2 \left( \beta - 4 b - 3 b' \right) \frac{\kappa^2 m}{8 \pi a r} + b H^2 \kappa^2 m \frac{\ln \left( \mathe^\gamma \bar{\mu} a r \right)}{4 \pi a r} + \frac{2 b}{3} \frac{\kappa^2 m}{4 \pi a^3 r^3} \eqend{.}
\end{equation}
Note that the terms containing $\ln a$, which came from the purely local part of the renormalized effective action $S_\text{loc,ren}$ of Eq.~\eqref{slocren_def} [resp. $S^{(\text{S})}_\text{loc,ren}$ of Eq.~\eqref{seff_bardeen}], have combined with the terms $\ln r$ coming from the non-local part $\Sigma_\text{ren}$ of Eq.~\eqref{sigmaren_def} [resp. $\Sigma^{(\text{S})}_\text{ren}$ of Eq.~\eqref{seff_bardeen2}].

\subsection{The flat-space limit}

The flat-space limit is given by taking $a = 1$ and $H = 0$. In this limit, all terms but the last disappear from the solutions~\eqref{sol_qu_a} and~\eqref{sol_qu_b}, and we obtain
\begin{equation}
\Phi_\text{H} = \Phi^\text{cl}_\text{H} + \kappa^2 \Phi^\text{qu}_\text{H} = \frac{\kappa^2 m}{16 \pi r} \left( 1 + \frac{4b}{3} \frac{\kappa^2}{r^2} \right)
\end{equation}
and
\begin{equation}
\Phi_\text{A} = \Phi^\text{cl}_\text{A} + \kappa^2 \Phi^\text{qu}_\text{A} = \frac{\kappa^2 m}{16 \pi r} \left( 1 + \frac{8b}{3} \frac{\kappa^2}{r^2} \right) \eqend{.}
\end{equation}
Since $\Phi_\text{A}$ gives the Newtonian potential $- V(r)$ in the Newtonian limit, let us give a more explicit expression for it for the case of free fields, where the coefficient $b$ is given in equation~\eqref{coeffs_b}. Substituting also $\kappa^2 = 16 \pi G_\text{N}$, we get
\begin{equation}
V(r) = - \frac{G_\text{N} m}{r} \left( 1 + \frac{N_0 + 6 N_{1/2} + 12 N_1}{45 \pi} \frac{G_\text{N}}{r^2} \right) \eqend{,}
\end{equation}
which is exactly the value computed in flat space by reconstructing the potential from scattering data and effective field equations~\cite{duff1974,capperduffhalpern1974,capperduff1974,duffliu2000a,duffliu2000b,parkwoodard2010,donoghue1994a,donoghue1994b,muzinichvokos1995,hamberliu1995,akhundovbelluccishiekh1997,kirilinkhriplovich2002,khriplovichkirilin2003,bjerrumbohrdonoghueholstein2003a,bjerrumbohrdonoghueholstein2003b,parkwoodard2010}.

\subsection{Comparison with previous results}

Wang and Woodard~\cite{wangwoodard2015} have calculated corrections to the gravitational potentials in de~Sitter space in the case of photons. Using the values of $b$ and $b'$ for this case~\eqref{coeffs_b}, we obtain
\begin{equation}
\label{sol_photon_a}
\Phi_\text{H} = \frac{\kappa^2 m}{16 \pi a r} \bigg[ 1 + \frac{\kappa^2}{120 \pi^2 a^2 r^2} + \frac{\kappa^2 H^2}{40 \pi^2} \left( \frac{19}{12} + 80 \pi^2 \beta + \ln \left( \mathe^\gamma \bar{\mu} a r \right) \right) \bigg]
\end{equation}
and
\begin{equation}
\label{sol_photon_b}
\Phi_\text{A} = \frac{\kappa^2 m}{16 \pi a r} \bigg[ 1 + \frac{\kappa^2}{60 \pi^2 a^2 r^2} + \frac{\kappa^2 H^2}{40 \pi^2} \left( \frac{7}{12} + 80 \pi^2 \beta + \ln \left( \mathe^\gamma \bar{\mu} a r \right) \right) \bigg] \eqend{.}
\end{equation}
We can then compare with the results of Ref.~\cite{wangwoodard2015}, using that they made a gauge choice such that their $h_{00}$ is equal to $\Phi_\text{A}$ and their $h_{ij}$ is equal to $\delta_{ij} \Phi_\text{H}$. While we agree on both the coefficient of the term which survives in the flat-space limit (in the first line) and the coefficient of the term involving $\ln r$, there are also some differences, which we list for clarity:
\begin{enumerate}
\item Wang and Woodard choose $\beta = 0$ for the arbitrary multiple of the square of the Ricci scalar.
\item The numerical coefficient of the quantum correction ($19/12$ for $\Phi_\text{H}$ and $7/12$ for $\Phi_\text{A}$) is equal to $1$ and $0$ in their result, respectively.
\item They choose the renormalization scale $\bar{\mu}$ to be equal to $\mathe^{-\gamma} H$.
\end{enumerate}
One physical difference to our calculation is the choice of the initial state, which Wang and Woodard take to be at $\eta_0 = 1/H$, while we take an adiabatic initial interaction vacuum state for $\eta_0 \to - \infty$. This difference in initial states may be responsible for the difference in finite parts of our results, but checking it explicitly is extremely involved and outside the scope of this work. However, one expects that corrections due to the initial state quickly redshift for late times, as $\eta \to 0$ ($a \to \infty$), which has been confirmed in concrete calculations~\cite{kahyaonemliwoodard2010}. In the late-time limit, the finite parts are subdominant with respect to the logarithmic term, and our calculations completely agree on this term.

Let us finally comment on an observation which gives us confidency in our result. The background de~Sitter metric in the Poincar\'e patch
\begin{equation}
\total s^2 = (-H\eta)^{-2} \left( - \total \eta^2 + \total \vec{x}^2 \right)
\end{equation}
is invariant under the rescaling
\begin{equation}
\label{rescaling}
\vec{x} \to \lambda \vec{x} \eqend{,} \qquad \eta \to \lambda \eta
\end{equation}
with constant $\lambda > 0$. Up to a global factor of $\lambda^{-2}$, this is also true for the effective field equations~\eqref{semiclassical_eqna} and~\eqref{semiclassical_eqnb}. The invariance is easily checked for all terms except the non-local one, which is invariant if the combination $H(x-x'; \bar{\mu}) + \delta^4(x-x') \ln a$ rescales with a factor of $\lambda^{-4}$. We thus calculate using the Fourier transform~\eqref{kernel_h_fourier}
\begin{equation}
\begin{split}
H(x-x'; \bar{\mu}) + \delta^4(x-x') \ln a &= \int \left[ \tilde{H}(\eta-\eta', \vec{p}; \bar{\mu}) + \delta(\eta-\eta') \ln a \right] \mathe^{\mathi \vec{p} (\vec{x}-\vec{x}')} \frac{\total^3 p}{(2\pi)^3} \\
&= \lim_{\epsilon \to 0} \int \bigg[ \cos\left[ \abs{\vec{p}} (\eta-\eta') \right] \frac{\Theta(\eta-\eta'-\epsilon)}{\eta-\eta'} \\
&\qquad\qquad+ \delta(\eta-\eta') \left( \ln(\bar{\mu} a \epsilon) + \gamma \right) \bigg] \mathe^{\mathi \vec{p} (\vec{x}-\vec{x}')} \frac{\total^3 p}{(2\pi)^3} \eqend{.}
\end{split}
\end{equation}
We now perform the rescaling~\eqref{rescaling}, as well as $\vec{p} \to \vec{p}/\lambda$ and $\epsilon \to \lambda \epsilon$. The combinations $\abs{\vec{p}} (\eta-\eta')$, $\vec{p} (\vec{x}-\vec{x}')$ and $a \epsilon$ then do not change, while
\begin{equation}
\frac{\Theta(\eta-\eta'-\epsilon)}{\eta-\eta'} \to \frac{\Theta[ \lambda(\eta-\eta'-\epsilon) ]}{\lambda(\eta-\eta')} = \lambda^{-1} \frac{\Theta(\eta-\eta'-\epsilon)}{\eta-\eta'}
\end{equation}
and
\begin{equation}
\delta(\eta-\eta') \to \delta[\lambda(\eta-\eta')] = \lambda^{-1} \delta(\eta-\eta') \eqend{.}
\end{equation}
The integration over $p$ furnishes another factor $\lambda^{-3}$, and since the limit $\lambda \epsilon \to 0$ is the same as $\epsilon \to 0$, we indeed obtain the required scaling
\begin{equation}
H[ \lambda(x-x'); \bar{\mu} ] + \delta^4[ \lambda(x-x') ] \ln \frac{a}{\lambda} = \lambda^{-4} \left[ H(x-x'; \bar{\mu}) + \delta^4(x-x') \ln a \right] \eqend{.}
\end{equation}
Our results~\eqref{sol_cl_ab},~\eqref{sol_qu_a} and~\eqref{sol_qu_b} only depend on the invariant combination $\hat{r} = a r = - r/(H \eta)$, and thus preserve this scale invariance.

\section{Discussion}
\label{sec_discussion}

We have calculated the correction to the gravitational potentials (the Bardeen variables) in de~Sitter space, due to the loop effects of conformal matter. Our results are valid for arbitrary conformal field theories, even strongly interacting ones, and depend on the parameters $b$ and $b'$ which appear in the trace anomaly (given by Eq.~\eqref{coeffs_b} for free theories, but generally taking different values for interacting theories). The result has the correct flat-space limit, and manifestly preserves the invariance under constant scaling of the coordinates, which is a symmetry of the underlying de~Sitter space. However, it differs from other recent calculations, and the cause of the discrepancy is currently unclear. An important difference with the correction to the Newtonian potential on a flat background is that here the leading quantum corrections also depend on the renormalized parameters appearing in front of the term quadratic in the curvature tensors of the gravitational action, which were needed to cancel the UV divergences due to the matter loops. In principle, one could thus measure these parameters by measuring the Newton potential, even if this is at present outside of experimental reach.

We can rewrite our result in a more suggestive form by reintroducing $\hbar$ and $c$, and expressing the result using the Planck length $\ell_\text{Pl} = \sqrt{\hbar G_\text{N}/c^3}$, the gravitational/Schwarzschild radius $\ell_\text{S} = 2 G_\text{N} m/c^2$ and the physical distance on equal-time hypersurfaces $\hat{r} \equiv a r$. Moreover, we consider arbitrary values of the renormalization scale $\mu$, and thus make the replacement~\eqref{barmu_replace}, which gives
\begin{equation}
\label{phi_a}
\Phi_\text{A} = \frac{\ell_\text{S}}{2 \hat{r}} \bigg[ 1 + \frac{128 \pi b}{3} \frac{\ell_\text{Pl}^2}{\hat{r}^2} + 32 \pi \ell_\text{Pl}^2 H^2 \left( \beta - 4 b - 3 b' + 2 c(\mu) + 2 b \ln \left( \mathe^\gamma \mu \hat{r} \right) \right) \bigg]
\end{equation}
and
\begin{equation}
\label{phi_h}
\Phi_\text{H} = \frac{\ell_\text{S}}{2 \hat{r}} \bigg[ 1 + \frac{64 \pi b}{3} \frac{\ell_\text{Pl}^2}{\hat{r}^2} + 32 \pi \ell_\text{Pl}^2 H^2 \left( \beta - 2 b - 3 b' + 2 c(\mu) + 2 b \ln \left( \mathe^\gamma \mu \hat{r} \right) \right) \bigg] \eqend{.}
\end{equation}
Just like the classical potentials~\eqref{sol_cl_ab}, also the quantum corrections are time-independent when expressed in terms of the physical distance $\hat{r}$. The terms in the first line are the ``de~Sitterized'' version of the known flat space results. Since they are of the form $(\ell_\text{Pl}/\hat{r})^2$, they are only important at very short distances, where the test particle approximation may break down. The terms in the second line are much more interesting. While the factor $\ell_\text{Pl}^2 H^2$ is extremely small at present times, during inflation it is small but appreciable, and it is conceivable that the logarithmic growth of the last term at large distances $\hat{r}$ could overcome this smallness and have potentially observable effects. However, this growth may be an artifact of perturbation theory. A well-known example is the infrared growth of loop corrections in massless, minimally coupled $\phi^4$ theory in de~Sitter space, which can be resummed using a variety of different methods~\cite{starobinsky1994,woodard2005,rajaraman2010,serreauparentani2013,youssefkreimer2014}. The non-perturbative result does not grow in the infrared, but shows strongly non-Gaussian correlations~\cite{starobinsky1994,woodard2005,riottosloth2008,rajaraman2010,benekemoch2013,serreauparentani2013,gautierserreau2013,youssefkreimer2014}. We thus find it prudent to say that our results~\eqref{phi_a} and~\eqref{phi_h} are valid as long as $\ell_\text{Pl}^2 H^2 \ln \hat{r} \ll 1$. To order $\ell_\text{Pl}^2$, to which we are working, we can combine the logarithm with the tree-level result into a modified power-law
\begin{equation}
\frac{\ell_\text{S}}{2 \hat{r}} \left[ 1 + 64 \pi b \ell_\text{Pl}^2 H^2 \ln \hat{r} \right] = \frac{\ell_\text{S}}{2 \hat{r}^{1 - 64 \pi b \ell_\text{Pl}^2 H^2}} + \bigo{\ell_\text{Pl}^4} \eqend{.}
\end{equation}
Thus, since $b > 0$ (which can be seen for free theories from Eq.~\eqref{coeffs_b}, and follows for general interacting theories from unitarity~\cite{osbornpetkou1994}) the potential decays slower at large distances, which we can interpret as an enhancement of the gravitational attraction due to quantum corrections. Furthermore, the effect depends on the Hubble scale, and is larger for larger $H$. Of course, there is also a constant correction of the overall strength which depends on the unknown coefficient $\beta$ and the renormalization-scale dependent $c(\mu)$, and this can be either negative or positive.

As was emphasized already previously, solving the effective field equations is a much simpler way to obtain the quantum corrections than reconstructing the potential from scattering data. In our opinion, the method presented in this work is even simpler than existing calculations using effective field equations~\cite{parkwoodard2010,wangwoodard2015,parkprokopecwoodard2015}. First, ``one effective action fits all'': while the calculation of the effective action involves little more than calculating the graviton self-energy (including renormalization), it can be used to study a variety of effects, and the effective action~\eqref{seff} has been used to study semiclassical stability of de~Sitter spacetime~\cite{fprv2013}, and to derive corrections to the tensor power spectrum~\cite{frv2012} and to the curvature tensor correlators~\cite{frv2014}. Second, since the effective action is gauge-invariant it can be expressed using only gauge-invariant variables, which provides a non-trivial check on the correctness of the calculations. This has two further advantages in the case treated here. On one hand, it reduces the number of differential equations from four to two (since two of the scalars appearing in the decomposition of the metric perturbations~\eqref{irreducible} are pure gauge), which shortens the calculation. On the other hand, this makes manifest that these equations are constraint equations, and that only spatial Laplacians need to be inverted [which is evident from the concrete expressions~\eqref{semiclassical_eqna}, \eqref{semiclassical_eqnb}]. Taken together, these facts considerably reduce the amount of calculations that need to be done: apart from purely algebraic manipulations, the only hard problem was the calculation of the non-local integral~\eqref{nonlocal_int}.

\begin{acknowledgments}
We would like to thank Richard Woodard, who encouraged us to look into this problem, and also Jaume Garriga, Shun-Pei Miao, Ian Morrison, Sohyun Park, Tomislav Prokopec, Albert Roura and Richard Woodard for fruitful discussions.

M.~F.\ acknowledges financial support through ERC starting grant QC\&C 259562. E.~V.\ acknowledges partial financial support from the Research Projects FPA2013-46570-C2-2-P, AGAUR 2014-SGR-1474, MDM-2014-0369 of ICCUB (Unidad de Excelencia `Mar{\'\i}a de Maeztu') and CPAN CSD2007-00042, with\-in the program Consolider-Ingenio 2010.
\end{acknowledgments}

\appendix

\section{Metric expansion}
\label{appendix_metric}

Given the perturbed metric $g_{\mu\nu}$ and expanding through quadratic order in the metric perturbation we have
\begin{equation}
\begin{split}
g_{\mu\nu} &= \eta_{\mu\nu} + h_{\mu\nu} \eqend{,} \\
g^{\mu\nu} &= \eta^{\mu\nu} - h^{\mu\nu} + h^\mu_\sigma h^{\nu\sigma} + \bigo{h^3} \eqend{,} \\
h &= \eta^{\mu\nu} h_{\mu\nu} \eqend{,} \\
\sqrt{-g} &= 1 + \frac{1}{2} h + \frac{1}{8} h^2 - \frac{1}{4} h_{\mu\nu} h^{\mu\nu} + \bigo{h^3} \eqend{,}
\end{split}
\end{equation}
where indices are raised and lowered with the unperturbed metric $\eta_{\mu\nu}$, i.e.\ we regard $h_{\mu\nu}$ as a tensor field in flat space. For the Christoffel symbols we get
\begin{equation}
\begin{split}
\Gamma^\alpha_{\mu\nu} &= \frac{1}{2} {S^\alpha}_{\mu\nu} - \frac{1}{2} h^\alpha_\sigma {S^\sigma}_{\mu\nu} + \bigo{h^3} \eqend{,} \\
{S^\alpha}_{\mu\nu} &\equiv \partial_\mu h^\alpha_\nu + \partial_\nu h^\alpha_\mu - \partial^\alpha h_{\mu\nu} \eqend{.} \\
\end{split}
\end{equation}
The calculation of the curvature tensors can be done straightforwardly and we obtain
\begin{equation}
\begin{split}
{R^\alpha}_{\beta\gamma\delta} &= \partial_{[\gamma} {S^\alpha}_{\delta]\beta} - h^\alpha_\sigma \partial_{[\gamma} {S^\sigma}_{\delta]\beta} - \frac{1}{2} \eta_{\mu\nu} \eta^{\alpha\sigma} {S^\mu}_{\sigma[\gamma} {S^\nu}_{\delta]\beta} + \bigo{h^3} \eqend{,} \\
R_{\alpha\beta} &= \frac{1}{2} \left( \partial_\mu {S^\mu}_{\alpha\beta} - \partial_\alpha \partial_\beta h \right) - h^\nu_\mu \partial_{[\nu} {S^\mu}_{\beta]\alpha} - \frac{1}{2} \eta_{\mu\nu} \eta^{\gamma\delta} {S^\nu}_{\delta[\gamma} {S^\mu}_{\beta]\alpha} + \bigo{h^3} \eqend{,} \\
R &= \left( \partial_\mu \partial_\nu h^{\mu\nu} - \partial^2 h \right) + h^{\mu\nu} \left( \partial_\nu \partial_\mu h + \partial^2 h_{\mu\nu} - 2 \partial_\nu \partial^\sigma h_{\mu\sigma} \right) \\
&\quad- \frac{1}{4} \left( 2 \partial_\sigma h^{\nu\sigma} - \partial^\nu h \right) \left( 2 \partial^\tau h_{\nu\tau} - \partial_\nu h \right) + \frac{1}{4} \left( 3 \partial_\gamma h_{\mu\delta} - 2 \partial_\mu h_{\gamma\delta} \right) \left( \partial^\gamma h^{\mu\delta} \right) + \bigo{h^3} \eqend{.}
\end{split}
\end{equation}
where $\partial^2 = \eta^{\mu\nu} \partial_\mu \partial_\nu$ and $\partial^\mu = \eta^{\mu\nu} \partial_\nu$.

The $n$-dimensional generalization of the Weyl tensor is given by
\begin{equation}
\label{weyl_tensor_def}
C^{\alpha\beta}{}_{\gamma\delta} = R^{\alpha\beta}{}_{\gamma\delta} - \frac{4}{(n-2)} R^{[\alpha}_{[\gamma} \delta_{\delta]}^{\beta]} + \frac{2}{(n-1)(n-2)} R \, \delta^\alpha_{[\gamma} \delta_{\delta]}^\beta \eqend{,}
\end{equation}
and the four-dimensional Euler density is given by
\begin{equation}
\label{euler_density_def}
\mathcal{E}_4 = R^{\alpha\beta\gamma\delta} R_{\alpha\beta\gamma\delta} - 4 R^{\alpha\beta} R_{\alpha\beta} + R^2 \eqend{.}
\end{equation}
For the square of the Weyl tensor we obtain the useful formula
\begin{equation}
C^{\alpha\beta\gamma\delta} C_{\alpha\beta\gamma\delta} = R^{\alpha\beta\gamma\delta} R_{\alpha\beta\gamma\delta} - \frac{4}{n-2} R^{\alpha\beta} R_{\alpha\beta} + \frac{2}{(n-1)(n-2)} R^2 \eqend{,}
\end{equation}
and thus the four-dimensional Euler density can be written also as
\begin{equation}
\mathcal{E}_4 = C^{\alpha\beta\gamma\delta} C_{\alpha\beta\gamma\delta} - 4 \frac{n-3}{n-2} R^{\alpha\beta} R_{\alpha\beta} + \frac{n (n-3)}{(n-1)(n-2)} R^2 \eqend{.}
\end{equation}

\section{Conformal transformation}
\label{appendix_conformal}

Under the conformal transformation
\begin{equation}
\tilde{g}_{\mu\nu} = a^2 g_{\mu\nu}
\end{equation}
the Christoffel symbols transform as
\begin{equation}
\tilde{\Gamma}^\alpha_{\mu\nu} = \Gamma^\alpha_{\mu\nu} + a^{-1} \left( \delta^\alpha_\mu \delta^\sigma_\nu + \delta^\alpha_\nu \delta^\sigma_\mu - g_{\mu\nu} g^{\alpha\sigma} \right) \partial_\sigma a \eqend{,}
\end{equation}
and the curvature tensors become
\begin{subequations}
\begin{align}
\tilde{R}^\alpha{}_{\beta\gamma\delta} &= R^\alpha{}_{\beta\gamma\delta} - 2 a^{-2} \delta^\alpha_{[\gamma} g_{\delta]\beta} ( \nabla a )^2 + 4 a^{-2} g^{\alpha\tau} \delta^\sigma_{[\gamma} g_{\delta][\tau} \left[ a \nabla_{\beta]} \nabla_\sigma a - 2 ( \nabla_{\beta]} a ) ( \nabla_\sigma a ) \right] \\
\begin{split}
\tilde{R}_{\mu\nu} &= R_{\mu\nu} - (n-2) a^{-1} \nabla_\mu \nabla_\nu a + 2 (n-2) a^{-2} ( \nabla_\mu a ) ( \nabla_\nu a ) \\
&\quad- g_{\mu\nu} \left[ (n-3) a^{-2} ( \nabla a )^2 + a^{-1} \nabla^2 a \right]
\end{split} \\
a^2 \tilde{R} &= R - (n-1) \left[ 2 a^{-1} \nabla^2 a + (n-4) a^{-2} ( \nabla a )^2 \right] \eqend{,}
\end{align}
\end{subequations}
where $\nabla_\mu$ is the covariant derivative associated with $g_{\mu\nu}$, $\nabla^2 = \nabla^\mu \nabla_\mu$, and where we used the notation
\begin{equation}
( \nabla a )^2 \equiv ( \nabla^\mu a ) ( \nabla_\mu a ) \eqend{.}
\end{equation}

The Weyl tensor~\eqref{weyl_tensor_def} does not change under a conformal transformation and we have
\begin{equation}
\label{weyl_conformal}
\tilde{C}^\alpha{}_{\beta\gamma\delta} = C^\alpha{}_{\beta\gamma\delta} \eqend{,}
\end{equation}
but for the Euler density we obtain
\begin{equation}
\label{euler_conformal}
\begin{split}
a^n \tilde{\mathcal{E}}_4 &= a^{n-4} \mathcal{E}_4 + \nabla_\mu \mathcal{E}^\mu - 4 (n-3) (n-4) G^{\mu\nu} a^{n-6} ( \nabla_\mu a ) ( \nabla_\nu a ) \\
&\quad+ (n-2) (n-3) (n-4) a^{n-8} ( \nabla a )^2 \left[ (5-n) ( \nabla a )^2 - 2 a \nabla^2 a \right]
\end{split}
\end{equation}
with
\begin{equation}
\begin{split}
\mathcal{E}^\mu &\equiv 4 (n-2) (n-3) a^{n-7} ( \nabla^\mu a ) \left( a \nabla^2 a + (n-4) ( \nabla a )^2 \right) \\
&\quad- 2 (n-2) (n-3) \nabla^\mu \left[ a^{n-6} ( \nabla a )^2 \right] + 8 (n-3) G^{\mu\nu} a^{n-5} \nabla_\nu a \eqend{,}
\end{split}
\end{equation}
using that $2 \nabla^\alpha G_{\alpha\beta} = 2 \nabla^\alpha R_{\alpha\beta} - \nabla_\beta R = 0$.

\section{Calculation of the non-local term}
\label{appendix_nonlocal}

In this appendix, we calculate the integral from equation~\eqref{nonlocal_int}. Performing a Fourier transform, we get
\begin{equation}
\begin{split}
I(x; \bar{\mu}) &= a(\eta) \int a^{-1}(\eta') \delta^3(\vec{x}') H(x-x'; \bar{\mu}) \total^4 x' = \int \left[ \int \frac{\eta'}{\eta} \tilde{H}(\eta-\eta', \vec{p}; \bar{\mu}) \total \eta' \right] \mathe^{\mathi \vec{p} \vec{x}} \frac{\total^3 p}{(2\pi)^3} \\
&\equiv \int \tilde{I}(\eta, \vec{p}; \bar{\mu}) \mathe^{\mathi \vec{p} \vec{x}} \frac{\total^3 p}{(2\pi)^3} \eqend{.} \raisetag{1.1\baselineskip}
\end{split}
\end{equation}
We now insert the Fourier transform~\eqref{kernel_h_fourier} of the kernel $H(x-x'; \bar{\mu})$, and include the $\mathi \epsilon$ prescription to select an interacting vacuum state as explained in subsection~\ref{sec_solutions_quantum}. In order not to confuse the two parameters $\epsilon$ (one coming from the proper definition of the distribution~\eqref{kernel_h_fourier} and one selecting the adiabatic interacting vacuum state), we denote the prescription parameter by $\delta$, and obtain
\begin{equation}
\begin{split}
\tilde{I}(\eta, \vec{p}; \bar{\mu}) = \lim_{\epsilon \to 0} \mathe^{ - \delta \abs{\vec{p}} \eta } \int &\mathe^{ \delta \abs{\vec{p}} \eta' } \frac{\eta'}{\eta} \cos\left[\abs{\vec{p}} (\eta-\eta')\right] \\
&\quad\times \left[ \frac{\Theta(\eta-\eta'-\epsilon)}{\eta-\eta'} + \delta(\eta-\eta') \left( \ln(\bar{\mu} \epsilon) + \gamma \right) \right] \total \eta' \eqend{,}
\end{split}
\end{equation}
with $\delta > 0$. The second part including the $\delta$ distribution is of course easily solved; for the other one we introduce an initial time $\eta_0$, express the cosine with exponentials and obtain
\begin{equation}
\tilde{I}(\eta, \vec{p}; \bar{\mu}) = \mathe^{ - \delta \abs{\vec{p}} \eta } \lim_{\epsilon \to 0} \left[ \Re \int_{\eta_0}^{\eta-\epsilon} \frac{\eta'}{\eta} \frac{\mathe^{\mathi \abs{\vec{p}} (\eta-\eta') + \delta \abs{\vec{p}} \eta'}}{\eta-\eta'} \total \eta' + \ln(\bar{\mu} \epsilon) + \gamma \right] \eqend{.}
\end{equation}
We then decompose
\begin{equation}
\frac{\eta'}{\eta} \frac{1}{\eta-\eta'} = \frac{1}{\eta-\eta'} - \frac{1}{\eta}
\end{equation}
and use the indefinite integral
\begin{equation}
\int \frac{\mathe^{a x}}{x-x_0} \total x = \mathe^{a x_0} \left[ \Ein[a(x-x_0)] + \ln(x-x_0) \right]
\end{equation}
to obtain
\begin{equation}
\begin{split}
\tilde{I}(\eta, \vec{p}; \bar{\mu}) &= \Re \left[ \Ein\left[ (\mathi-\delta) \abs{\vec{p}} (\eta-\eta_0) \right] + \ln(\eta-\eta_0) - \frac{\mathe^{- (\mathi-\delta) \abs{\vec{p}} (\eta_0-\eta)}}{\eta (\mathi-\delta) \abs{\vec{p}}} \right] \\
&\quad+ \lim_{\epsilon \to 0} \left[ \Re \left[ - \Ein\left[ (\mathi-\delta) \abs{\vec{p}} \epsilon \right] - \ln(\epsilon) + \frac{\mathe^{(\mathi-\delta) \abs{\vec{p}} \epsilon}}{\eta (\mathi-\delta) \abs{\vec{p}}} \right] + \mathe^{ - \delta \abs{\vec{p}} \eta } \left( \ln(\bar{\mu} \epsilon) + \gamma \right) \right] \eqend{.}
\end{split}
\end{equation}
The terms in the first line depend on $\eta_0$ and must be absorbed in initial state corrections if $\delta = 0$. We choose the $\mathi \delta$ prescription to select an interacting vacuum state, and thus take first $\eta_0 \to -\infty$ and then $\delta \to 0$. Since
\begin{equation}
\begin{split}
\Ein\left[ (\mathi-\delta) \abs{\vec{p}} (\eta-\eta_0) \right] &\sim - \gamma - \ln\left[ - (\mathi-\delta) \abs{\vec{p}} (\eta-\eta_0) \right] \\
&\qquad+ \mathe^{(\mathi-\delta) \abs{\vec{p}} (\eta-\eta_0)} \left[ \frac{1}{- (\mathi-\delta) \abs{\vec{p}} \eta_0} + \bigo{\eta_0^{-2}} \right] \eqend{,}
\end{split}
\end{equation}
the first line gives taking first the limit $\eta_0 \to -\infty$ holding $\delta > 0$ fixed, and then $\delta \to 0$
\begin{equation}
- \gamma - \ln \abs{\vec{p}} \eqend{.}
\end{equation}
For small argument, $\Ein$ vanishes, and thus the limits $\epsilon \to 0$ and $\delta \to 0$ can be easily taken for the second line. Summing the terms from both lines, in total we obtain the extremely simple result
\begin{equation}
\tilde{I}(\eta, \vec{p}; \bar{\mu}) = \ln \frac{\bar{\mu}}{\abs{\vec{p}}} \eqend{.}
\end{equation}

It remains to perform the inverse Fourier transform. For this, it is convenient to calculate the integral
\begin{equation}
J(x) \equiv \int \abs{\vec{p}}^{-2} \ln\abs{\vec{p}} \mathe^{\mathi \vec{p} \vec{x}} \frac{\total^3 p}{(2\pi)^3} = \frac{1}{2 \pi^2} \lim_{\epsilon \to 0} \int_0^\infty \mathe^{-\epsilon p} \ln p \frac{\sin(p r)}{p r} \total p
\end{equation}
with $r = \abs{\vec{x}}$, and then recover the needed inverse Fourier transform in the form
\begin{equation}
\label{i_from_j}
I(x; \bar{\mu}) = \int \ln \frac{\bar{\mu}}{\abs{\vec{p}}} \mathe^{\mathi \vec{p} \vec{x}} \frac{\total^3 p}{(2\pi)^3} = \ln \bar{\mu} \, \delta^3(\vec{x}) + \laplace J(x) \eqend{.}
\end{equation}
We first integrate by parts once to obtain
\begin{equation}
\begin{split}
\int_0^\infty \mathe^{-\epsilon p} \ln p \frac{\sin(p r)}{p r} \total p &= \frac{1}{2r} \int_0^\infty \mathe^{-\epsilon p} \sin(p r) \frac{\partial \ln^2 p}{\partial p} \total p \\
&= - \frac{1}{2r} \int_0^\infty \mathe^{-\epsilon p} [ r \cos(p r) - \epsilon \sin(p r) ] \ln^2 p \total p \\
&= \Im \left[ \frac{\epsilon-\mathi r}{2r} \int_0^\infty \mathe^{- (\epsilon-\mathi r) p} \ln^2 p \total p \right] \eqend{.}
\end{split}
\end{equation}
We then write
\begin{equation}
\ln^2 p = \lim_{h \to 0} \left( \frac{p^h - 1}{h} \right)^2 = \lim_{h \to 0} \frac{p^{2h} - 2 p^h + 1}{h^2}
\end{equation}
and use that for $\Re a < 0$ and $b \geq 0$ we have
\begin{equation}
\int_0^\infty \mathe^{a p} p^b \total p = \frac{\Gamma(b+1)}{(-a)^{b+1}}
\end{equation}
to obtain
\begin{equation}
\int_0^\infty \mathe^{- (\epsilon-\mathi r) p} \ln^2 p \total p = \lim_{h \to 0} \frac{1}{h^2} \left[ \frac{\Gamma(1+2h)}{(\epsilon-\mathi r)^{1+2h}} - 2 \frac{\Gamma(1+h)}{(\epsilon-\mathi r)^{1+h}} + \frac{1}{\epsilon-\mathi r} \right] \eqend{.}
\end{equation}
The $\Gamma$ function expansion
\begin{equation}
\Gamma(1+\alpha h) = 1 - \gamma \alpha h + \frac{6 \gamma^2 + \pi^2}{12} \alpha^2 h^2 + \bigo{h^3}
\end{equation}
then gives
\begin{equation}
\int_0^\infty \mathe^{- (\epsilon-\mathi r) p} \ln^2 p \total p = \frac{\left[ \gamma + \ln (\epsilon-\mathi r) \right]^2 + \pi^2/6}{\epsilon-\mathi r}
\end{equation}
and thus
\begin{equation}
\int_0^\infty \mathe^{-\epsilon p} \ln p \frac{\sin(p r)}{p r} \total p = \frac{1}{2r} \Im \left[ \gamma + \ln (\epsilon-\mathi r) \right]^2 \eqend{.}
\end{equation}
We can now take the limit $\epsilon \to 0$ to get
\begin{equation}
\ln (\epsilon-\mathi r) = \ln r - \frac{\mathi \pi}{2} + \bigo{\epsilon}
\end{equation}
and
\begin{equation}
J(x) = - \frac{\gamma + \ln r}{4 \pi r} \eqend{.}
\end{equation}
Note that this is a well-defined distribution in three spatial dimensions. Using the well-known identity
\begin{equation}
\laplace \frac{1}{r} = - 4 \pi \delta^3(\vec{x}) \eqend{,}
\end{equation}
we then obtain from equation~\eqref{i_from_j}
\begin{equation}
\label{appendix_nonlocal_i}
I(x; \bar{\mu}) = - \laplace \frac{\ln \left( \mathe^\gamma \bar{\mu} r \right)}{4 \pi r} \eqend{,}
\end{equation}
which also is a well-defined distribution.

\bibliography{literature}

\end{document}